\documentclass[superscriptaddress,aps,preprintnumbers,amsmath,amssymb,nofootinbib]{revtex4}
\usepackage{graphicx}
\usepackage{epstopdf} 
\usepackage{dcolumn}% Align table columns on decimal point
\usepackage{bm}% bold math

\def\slashchar#1{\setbox0=\hbox{$#1$} % set a box for #1
\dimen0=\wd0 % and get its size
\setbox1=\hbox{/} \dimen1=\wd1 % get size of /
\ifdim\dimen0>\dimen1 % #1 is bigger
\rlap{\hbox to \dimen0{\hfil/\hfil}} % so center / in box
#1 % and print #1
\else % / is bigger
\rlap{\hbox to \dimen1{\hfil$#1$\hfil}} % so center #1
/ % and print /
\fi}

\def\a{\alpha}
\def\b{\beta}

\def\g{\gamma}

\def\k{\kappa}
\def\l{\lambda}
\def\m{\mu}
\def\n{\nu}

\def\r{\rho}
\def\s{\sigma}

\def\u{\upsilon}

\def\x{\xi}

\def\D{\Delta}
\def\G{\Gamma}

\def\beq{\begin{eqnarray}}
\def\eeq{\end{eqnarray}}

\newcommand{\vev}[1]{ \left\langle {#1} \right\rangle }

\begin{document}
\title{R-axion detection at LHC}

\author{Hock-Seng Goh}
\affiliation{%
Department of Physics, University of California, Berkeley,
and Theoretical Physics Group, LBNL, Berkeley, CA 94720
}%
%\altaffiliation[Also at ]{Physics Department, XYZ University.}
% \altaffiliation[Also at ]{Physics Department, XYZ University.}
\author{Masahiro Ibe}
\affiliation{%
Stanford Linear Accelerator Center, Stanford University, Stanford, CA 94309
%Physics Department, Stanford University, Stanford, CA 94305
}%
%\altaffiliation[Also at ]{Physics Department, XYZ University.}
% \altaffiliation[Also at ]{Physics Department, XYZ University.}
%\altaffiliation[Also at ]{Physics Department, XYZ University.}
% \altaffiliation[Also at ]{Physics Department, XYZ University.}
%\affiliation{}
%\affiliation{%
%\affiliation{

\begin{abstract}
Supersymmetric models with spontaneously broken approximate R-symmetry contain a light spin
$0$ particle, the R-axion.
The properties of the particle can be a powerful probe of the structure of the new physics.
In this paper, we discuss the possibilities of the R-axion detection at the LHC experiments.
It is challenge to observe this light particle in the LHC environment.
However, for typical values in which the mass of the R-axion is a few hundred MeV,
we show that those particles can be detected
by searching for displaced vertices from R-axion decay.
\end{abstract}

\date{\today}
\maketitle
\preprint{SLAC-PUB-13451}
\section{Introduction}
% Generic Introduction to the R-symmetry and its breaking
Discovering supersymmetry is one of the main goal of the LHC experiments.
Once the supersymmetric standard model (SSM) is confirmed experimentally,
the next question is how the supersymmetry is broken and how the effects of symmetry breaking
are mediated to the SSM sector.
In most cases, such investigations on ``beyond the SSM physics'' rely on arguments based
on extrapolations of the observed supersymmetry mass parameters to higher energies.
However, there is one class of models of supersymmetry breaking  in which  we can get a glimpse
of the structure of the hidden sector in a direct way with the help of the R-symmetry.

The R-symmetry is a symmetry which is rather generic in the models of spontaneous
supersymmetry breaking through $F$-term\,\cite{Nelson:1993nf}.
At the same time, however, it must be broken in some way in order
for  the gauginos in the SSM sector to have non-vanishing masses,
which we are assuming to be the Majorana fermions.
One possibility of the gaugino mass generation is to consider  models
where the gaugino masses are generated as a result of the explicit breaking of the R-symmetries,
which is realized in the supergravity mediation scenario\,\cite{Chamseddine:1982jx,Hall:1983iz},
the anomaly mediation scenarios\,\cite{Randall:1998uk},
and in a class of the gauge mediation\,\cite{Dine:1981za,Dine:1993yw,Dine:1994vc,Dine:1995ag}
with the explicit R-symmetry
breaking\,\cite{Kitano:2006wm,Murayama:2006yf,Ibe:2006rc,Kitano:2006xg,Nomura:2007cc,Haba:2007rj}.
Unfortunately, in those models, the R-symmetry leaves little trace for the collider experiments,
since the mass of the R-axion is typically heavy and beyond the reach of the LHC experiments.

In this paper, instead, we focus on a class of  models with gauge mediation where the
R-symmetry in the hidden/messenger sectors is exact in the limit of the infinite reduced Planck scale,
i.e. $M_{\rm PL}\to \infty$.
Then, the gaugino masses are generated as a result of the spontaneous breaking of the R-symmetry.
We also assume that the R-symmetry is respected by the SSM sector
as well as the sector of the origin of the higgsino mass $\mu$ and the Higgs mass mixing $B\mu$
except for the anomaly to the SSM gauge interactions.
We call this scenario, the minimal R-symmetry breaking scenario.

The immediate consequence of the spontaneously broken approximate R-symmetry is
the existence of a pseudo Nambu-Goldstone boson, so called the R-axion\,\cite{Nelson:1993nf}.
As is well know, the R-axion can have a large mass\,\cite{Bagger:1994hh}
compared with the Peccei-Quinn axion\,\cite{Peccei:1977hh,Weinberg:1977ma},
while getting small couplings.
Hence, it is free from cosmological and astrophysical constraints.
The notable feature of the R-axion in the minimal R-symmetry breaking scenario is that
it couples to the gluinos and this enhances the R-axion-gluon coupling at the low energy
effective theory.
As we will show, a sizable number of R-axions can be produced via the gluon fusion
process with the dominant contribution coming from
gluino and messenger loops.
Since the R-axion has long lifetime, it is possible to detect the R-axion
by searching for the displaced vertex of the R-axion decay.

It is possible that the R-axion can produce the following striking signature:
A promptly produced jet of $p_{T}>100$\,GeV, recoiling against a low mass muon
pair produced in the beam pipe vacuum region.
In the paper, we will compute the rate of background for this signature.

The organization of the paper is as follows.
In the next section, we briefly review the generic properties of R-symmetry breaking
in models with gauge mediation.
In section\,\ref{sec:rint}, we summarize interactions of the R-axion with the SSM particles.
Now the paper shifts from a theoretical
to an experimental viewpoint.
In section\,\ref{sec:production}, we discuss the production of the R-axion and the typical signatures.
In section\,\ref{sec:background},
we give a rough estimation of the background processes to the R-axion signal.
The final section gives our conclusions.

%%%%%%%%%%%%%%%%%%%%%%%%%%%%%%%%
\section{R-symmetry in Gauge Mediation Scenario}\label{sec:rsym}
As examples of models of the R-symmetry breaking,
we consider two simple examples of the models with gauge mediation.
The first example is the so-called `minimal gauge mediation' model
in which the supersymmetry breaking effects  are mediated by
a pair messenger fields $\psi$ and $\bar{\psi}$
which are charged under the SSM gauge group.
To handle the supersymmetry breaking, we introduce a chiral superfield $S$ whose  $F$-term
has a non-vanishing vacuum expectation value (VEV) called $F$ in the following discussion.
In the minimal model, the messenger fields and the supersymmetry breaking field couple
through the superpotential,
\begin{eqnarray}
 W = k S \bar\psi \psi,
\end{eqnarray}
where $k$ is a coupling constant.
The above superpotential possesses an R-symmetry with a charge assignment $S(2)$,
$\psi(0)$, and $\bar{\psi}(0)$,
which forbids the gaugino mass if it is not broken.
This charge assignment is consistent with $F\neq 0$, since it implies that the  $F$-term
has R-charge 0; other possible $U(1)$'s, consistent with nonzero gaugino masses, are
broken by $F$.

In this example, the scalar VEV of the chiral superfield $S$ provides a supersymmetric mass term
to the messenger particles, which corresponds to the order parameter of the R-symmetry.
Then, in order that the messenger fields are not tachyonic,
the scalar and the $F$-term VEVs of the chiral superfield $S$,
\begin{eqnarray}
 \vev{S} = M + F \theta^{2},\quad
\end{eqnarray}
are required to satisfy the condition,
\begin{eqnarray}
|kM| > |\sqrt F |.
\end{eqnarray}
Thus, in the minimal gauge mediation,  the R-symmetry must be broken, with a VEV even larger
than the supersymmetry breaking scale as long as $k$ is perturbative.

Another example is a model proposed in Ref.\,\cite{Izawa:1997gs}.
Let us add another pair of the messenger particles $\psi'$ and $\bar\psi'$ so that we can give the supersymmetric and R-symmetric mass terms to the original messengers;
\begin{eqnarray}
 W = k S \psi\bar\psi + m \psi \bar\psi' +  m \psi' \bar\psi,
\end{eqnarray}
where $m$ denotes a mass parameter. Here, again, the R-symmetry
which is relevant for the gaugino mass suppression is defined by
$S(2)$, $\psi(0)$, $\bar{\psi}(0)$, $\psi'(2)$, and
$\bar{\psi}'(2)$. In this model, the messenger fields are not
tachyonic even if $M = 0$ contrary to the former example. However,
even in this case, there is a lower bound on the size of $M$
coming from the experimental lower bound on the gaugino masses.
That is, in this model, the gaugino mass is roughly given by,
\begin{eqnarray}
 m_{\rm gaugino} \sim \frac{\a}{4\pi}\frac{k M}{m} \left|\frac{kF}{m^{2}} \right|^{2}\frac{kF^{*}}{m}.
\end{eqnarray}
Then, remembering that $|kF/m^{2}| < 1$ so that the messengers are not tachyonic,
we obtain a lower bound on $M$,
\begin{eqnarray}
M > \frac{4\pi}{k\a}m_{\rm gaugino},
\end{eqnarray}
where the lower bound is saturated for $m \simeq \sqrt{F}$.

The above discussion applies to general gauge mediation models with R-symmetric
hidden/messenger sectors.
In a general gauge mediation model, the gaugino masses can be parameterized by,
\begin{eqnarray}
m_{\rm gaugino} \simeq N_{m}\frac{\a}{4\pi}\frac{M}{M_{m}}  \frac{F^{*}}{M_m}
\left(C_{m} + O\left( \frac{F}{M_m^{2}} \right)\right),
\end{eqnarray}
where  the coefficient $C_{m}$ is a model dependent parameter at most  $O(1)$, $M_{m}(\gtrsim M)$
is the scale of the gauge mediation, and $N_{m}$ is the number of the messenger fields.%
\footnote{For recent developments on the analysis of the models with the general gauge mediation,
see Refs.\,\cite{Meade:2008wd,Carpenter:2008wi,Ooguri:2008ez,Distler:2008bt,Intriligator:2008fr}.}
Here, we have omitted $O(1)$ coupling constants of the mediation mechanism.
From this expression, we obtain a lower bound on the R-symmetry breaking scale,
\begin{eqnarray}
\label{eq:fm}
M \gtrsim 10^{4}\,{\rm GeV} \left(\frac{5}{N_{m}} \right)
\left(\frac{m_{\rm gluino}}{500\,\rm GeV} \right),
\end{eqnarray}
which results from the non-tachyonic messenger condition,  $M_{m} \gtrsim \sqrt{F}$.%
\footnote{The number of the messenger fields is consistent with
the perturbative grand unification theory (GUT) as long as it satisfies
a condition, $N_{m}\lesssim 150/\ln(M_{\rm GUT}/M_{\rm mess})$.}
The lower bound is saturated for the lowest scale gauge mediation, i.e. $M_{m} \simeq \sqrt{F}$.
Therefore, we see that the R-symmetry breaking scale is about $10^{4}$\,GeV
or higher in generic gauge mediation models.

Notice that the masses of the SSM scalars are roughly given by,
\begin{eqnarray}
 m_{\rm scalar} \sim \sqrt{N_{m}}\frac{\a}{4\pi}\frac{F}{M_{m}},
\end{eqnarray}
and that there is no suppression from the R-symmetry breaking.
Thus, when the R-symmetry breaking scale is suppressed compared with $M_{m}$ and $\sqrt{F}$,
the gaugino masses are suppressed compared with the scalar masses.
In other words,
the SSM scalars are much heavier than the electroweak scale even if the gauginos have the masses
of the order of the electroweak scale.
However, such heavy scalars bring up the hierarchy problem again;
we do not pursue the possibility of such a highly hierarchical spectrum in this study.
Instead, we consider models where the gaugino and scalar masses
are not so different, i.e. $M\sim M_{m}\sim \sqrt{F}$,
which is naturally realized in low scale direct gauge mediation models
(see for example\,\cite{Izawa:1997hu,Izawa:2005yf,Ibe:2007wp,Ibe:2007ab}).

In the following, we confine ourselves to the scenario with the low R-symmetry breaking
scale as well as the low scale supersymmetry breaking and the messenger scale,
i.e the lowest gauge mediation scenario;
\begin{eqnarray}
\label{eq:lowscale}
 f_{R} &=& 10^{4-5}\, {\rm GeV},\cr
 \sqrt{F} &=& 10^{4-5}\, {\rm GeV},\cr
 M_{m} &=& 10^{4-5}\,{\rm GeV},
\end{eqnarray}
where $f_{R}$ denotes the R-symmetry breaking scale.
As we will show in the subsequent sections, a sizable number of R-axions can be produced
in the LHC experiments for such low scale gauge mediation models.

We mention here that the low scale gauge mediation scenario is also motivated from cosmology.
The low scale gauge mediation scenario given above admits a very light gravitino
with the mass lighter than $O(10)$\,eV.
Such a light gravitino is free from all the cosmological problems\,\cite{Viel:2005qj}.
The model has the phenomenology typical in gauge mediation
that the SSM particle is unstable and decays to the gravitino.

%%%%%%%%%%%%%%%%%%%%%%%%%%%%%%%%
\subsection*{R-axion Mass}
One immediate consequence of the spontaneous breaking of the R-symmetry breaking
is the R-axion.
In our scenario, we have assumed that,
if supergravity in the action is ignored ($M_{\rm PL}\to \infty$),
the R-symmetry is only broken by the anomalies of the SSM gauge interactions.
In this limit, the R-axion obtains a mass only from the QCD anomaly.
However, supergravity adds several R-symmetry breaking effects
which contribute to the mass of the R-axion\,\cite{Nelson:1993nf}.
In particular,  the R-symmetry is broken by the constant term in the superpotential
which is always required to set the cosmological constant  to zero after the supersymmetry
breaking.%
\footnote{We may further postulate that the constant term also emerges as a result
of the spontaneous symmetry breaking of another R-symmetry.
In this case, we have an additional axion.
The additional axion can provide the solution to the strong CP-problem.}
For example, in the case of the simplest supersymmetry breaking sector model
with the full superpotential,
\begin{eqnarray}
W = \Lambda_{\rm susy}^{2} S + w\,,
\end{eqnarray}
the R-symmetry is explicitly  broken by the constant term $w$.
Here $\Lambda_{\rm susy}$ denotes a dimensionful constant such that
the $F$-term VEV of $S$ is given by $\vev F=\Lambda_{\rm susy}^{2}$,
and the flat universe condition requires $\Lambda_{\rm susy}^{4}=3 w^{2}/M_{\rm PL}^{2}$.
In the supergravity, this superpotential leads to
the R-symmetry breaking term in the scalar potential of $S$,
\begin{eqnarray}
V_{\slashchar  R} =-2 \frac{w}{M_{P}^{2}} \Lambda_{\rm susy}^{2} S + c.c.
=-2 m_{3/2} \Lambda_{\rm susy}^{2} S + c.c.
\end{eqnarray}
Here we have used the definition of the gravitino mass $m_{3/2} = w/M_{P}^{2}$.
$S$ obtains a VEV, and we parametrize this by,  $S = f_{R}/\sqrt{2}\, e^{i \tilde{a}/f_{R}}$,
$f_{R}$ then gives the scale of spontaneous R-symmetry breaking.
As a result of the breaking term $V_{\slashchar R}$, the R-axion obtains a non-vanishing mass
\begin{eqnarray}
 m_{a}^{2} = \frac{2\sqrt{2}m_{3/2} \Lambda_{\rm susy}^{2}}{f_{R}}
 = \frac{2\sqrt{6}m_{3/2}^{2} M_{\rm PL}}{f_{R}} \,.
\end{eqnarray}
Thus, if we consider the low breaking scale and the light gravitino mass, the R-axion mass is given by,
\begin{eqnarray}
\label{eq:Rmass}
 m_{a} = 343\,{\rm MeV}
 \left(\frac{m_{3/2}}{10\, \rm eV} \right) \left(\frac{f_{R}}{10^{4}\, \rm GeV} \right)^{-1/2}\,.
\end{eqnarray}

In general models, we could consider other explicit R-symmetry breaking terms that  are also
vanishing in the limit (see Ref.\,\cite{Bagger:1994hh} for more discussion).
Those breaking terms also contribute to the R-axion mass at the similar size of the
above contribution especially for the lowest gauge mediation scenario.
We should also notice that, in general models, there is a distribution between
the total supersymmetry breaking scale, $\Lambda_{\rm susy}$, which determines
the gravitino mass, and the scale used in the gauge meditation, $\sqrt{F}$.
Therefore, the R-axion mass depends on the details of the models of the hidden/messenger sectors
and  could be larger than the value given in Eq.\,(\ref{eq:Rmass}).%
\footnote{In models in \cite{Izawa:1997gs,Fujii:2003iw}, the R-axion can be
much lighter than the values discussed above, thanks to the scale invariance
of the superpotential of the supersymmetry breaking field.
See also Refs.\,\cite{Demir:2001ky} for related works.
 }

In the following analysis,
we take the R-axion as an independent parameter.
We mainly focus on the range of  R-axion masses,
\begin{eqnarray}
 m_{a} = O(100)\,{\rm MeV},
 \end{eqnarray}
 which is, for example, the most likely mass range
 for $f_{R}\sim \sqrt{F}\gtrsim 10^{4}$\,GeV and a light gravitino mass,
 $m_{3/2} = O(10)$\,eV in the above simple supersymmetry breaking scenario.
 As we will see in the next section,
R-axions in this range have long lifetimes and leave displaced vertices inside the detecters.

%%%%%%%%%%%%%%%%%%%%%%%%%%%%%%%%
\section{Interactions of R-axion}\label{sec:rint}
Let us now consider the interactions between the R-axion and the SSM sector.
Here, we consider the minimal Higgs sector model with two Higgs doublets (i.e. the MSSM),
although we can easily extend our analysis to models with a non-minimal Higgs sector.
As we will see, the R-axion mixes with the pseudoscalar components in the two Higgs doublets
at the energy scale below the electroweak symmetry breaking.%
\footnote{In this study, we assume that there is no supersymmetric CP-problem,
which is one of the motivation of the models with the gauge mediation.
}
The mixing angles determine the couplings between the R-axion and the
SSM matter fields.

For our purpose, the effective Lagrangian approach valid below the messenger scale
is the most useful.
After integrating out the hidden/messenger sectors, the low energy effective theory is given by the SSM particles and the R-axion $\tilde{a}$.
In the effective theory, the R-symmetry is realized by the shift symmetry of the R-axion,
\begin{eqnarray}
 \tilde{a}/f_{R} &\to& \tilde{a}'/f_{R} = \tilde{a}/f_{R} + 2c,
\end{eqnarray}
with the angle $c$, as well as by the linear transformation of the SSM fields with the charges given in Table~\ref{tab:R}.
Notice that the R-charges of the SSM fields satisfy,
\begin{eqnarray}
\label{eq:charge}
X_{u} + X_{d} = 2, \quad
X_{u} + X_{Q} + X_{\bar U} = 2, \quad
X_{d} + X_{Q} + X_{\bar D} = 2,
\end{eqnarray}
so that the R-symmetry is consistent with the $\m$-term and the Yukawa interactions.%
\footnote{
In this study, we assume that the $\m$-term carries a vanishing R-charge.
In models where the $\m$-term is also generated via the interactions
between the hidden/messenger sectors and the Higgs sector, however,
the resultant $\m$-term can have non-vanishing R-charge.
The following analysis can be generalized straightforwardly to such models.
}

In the effective theory just below the messenger scale, the R-axion only emerges in the gaugino mass
terms, and the anomaly coupling terms.
The effective Lagrangian of the R-axion is given by,
\begin{eqnarray}
\label{eq:RaxionL}
{\cal L}_{\tilde a} =  \frac{1}{2} (\partial \tilde a)^{2} - m_{1/2}^{i} e^{-i {\tilde a}/{f_{R}}}
\lambda^{i} \lambda^{i}+ C_{H}\frac{g_{i}^{2}}{32\pi^{2}}
\frac{\tilde{a}}{f_{R}}
F^{i}\tilde{F}^{i}
+ c.c.,
\end{eqnarray}
where $g_{i}$ denotes the gauge coupling constant,  $m_{1/2}^{i}$ the gaugino mass,
$\tilde {F}_{\m\n} = \epsilon_{\m\u\r\s} F^{\r\s}/2$, and $i$ runs over the SSM gauge groups.
The coefficient of the anomaly coupling terms $C_{H}$ depends
on the model of the hidden/messenger sectors.
From the second term, we find that the R-axion couples to the gauginos by a Yukawa-type interaction,
\begin{eqnarray}
 {\cal L} _{\rm int} = - i \frac{m_{1/2}^{i}}{f_{R}}\tilde{a}\lambda^{i} \lambda^{i} + c.c.
\end{eqnarray}

%%%%%%%%%%%%%%%%%%%%%%%%%%%%%%%%%%%%%%%%%%%%%
\begin{table}[tb]
\begin{center}
\begin{tabular}{c|c c c}
& SU(2) & U(1)$_{\rm Y}$ & U(1)$_{\rm R}$\\
\hline
$H_{u}$ & ${\bf 2}$ & $1/2$  & $X_{u}$ \\
$H_{d}$ & ${\bf 2}$ & $-1/2$ & $X_{d}$ \\
$Q_{L}$ & ${\bf 2}$ & $1/6$ & $X_{Q}$ \\
$\bar{U}_{R}$ & ${\bf 1}$ & $-2/3$ & $X_{\bar U}$ \\
$\bar{D}_{R}$ & ${\bf 1}$ & $1/3$ & $X_{\bar D}$ \\
$\lambda$ & - & - & $1$
\end{tabular}
\end{center}
\caption{The R-charges of the supermultiplets in the SSM and gauginos.
Here, we also showed the charges of them under the Standard Model gauge group; SU(2)$\times$U(1)$_{Y}$. The charge assignment is identical for all the three generations. }
\label{tab:R}
\end{table}%
%%%%%%%%%%%%%%%%%%%%%%%%%%%%%%%%%%%%%%%%%%%%%

At a lower energy scale, the R-axion dependence in the gaugino masses spreads to the other soft supersymmetry breaking masses, such as the $B\m$ and the $A$-terms,
as a result of  renormalization group evolution.
The R-axion dependence in the $B\m$-term plays a particularly important role
in determining the mixing between the R-axion and the SSM Higgs bosons.
The relevant terms for the R-axion-Higgs mixing are given by,
\begin{eqnarray}
 V = (|\m|^{2} + m_{H_{u}}^{2} ) |H_{u}^{2}| +
 (|\m|^{2} + m_{H_{d}}^{2} ) |H_{d}^{2}|
 - (  e^{-i\tilde a/f_{R}} B\m H_{u}^{0} H_{d}^{0} + c.c.)
+ \frac{1}{8}(g^{2}+{g'}^{2}) (|H_{u}^{0}|^{2} - |H_{d}^{0}|^{2} )^{2},
\end{eqnarray}
where $H_{u,d}^{0}$ are the QED neutral components in the Higgs doublets.
Notice that in these models, the $B\m$ term is generated by the renormalization group evolution
coming from gaugino loop diagrams\,\cite{Giudice:1997ni} and so is proportional to $e^{-i \tilde{a}/f_{R}}$.%
\footnote{The $B\m$ term also gets small contribution from the two-loop diagrams
in which the messengers circulate as well as the gauginos and Higgsinos\,\cite{Giudice:1997ni},
which is also proportional to $e^{-i \tilde{a}/f_{R}}$.
}
To analyze the mixing mass matrix, it is convenient to
decompose the neutral Higgs fields into the radial and the axial components,
\begin{eqnarray}
 H_{u}^{0} = \frac{1}{\sqrt 2} (v_{u}+ \rho_{u}) e^{i \xi_{u}/v_{u}},\quad
 H_{d}^{0} = \frac{1}{\sqrt 2} (v_{d}+ \rho_{d}) e^{i \xi_{d}/v_{d}}.
\end{eqnarray}
Here, $v_{u,d}$ are vacuum expectation values of the Higgs fields which
are related to the $Z^{0}$ boson mass by
\begin{eqnarray}
 v^{2}\equiv v_{u}^{2} + v_{d}^{2} = 4 m_{Z}^{2}/(g^{2}+{g'}^{2})\simeq (246\, {\rm GeV})^{2}.
\end{eqnarray}
The ratio between $v_{u}$ and $v_{d}$ is traditionally expressed by,
\begin{eqnarray}
 \tan\beta \equiv v_{u}/v_{d}.
\end{eqnarray}
Then, with the aid of the minimum conditions of the $H_{u}^{0}$ and $H_{d}^{0}$;
\begin{eqnarray}
|\m|^{2} + m_{H_{u}}^{2} &=& B\m \cot \b + (m_{Z}^{2}/2) \cos2\b,\cr
|\m|^{2} + m_{H_{d}}^{2} &=& B\m \tan \b  - (m_{Z}^{2}/2) \cos2\b,
\end{eqnarray}
we obtain the mixing mass matrix of the axial components and the R-axion,
\begin{eqnarray}
V_{\rm mix} = \frac{1}{2} {\bf x}^{t} {\cal M}^{2} {\bf x},\quad
{\bf x }=
\left(
\begin{array}{c}
 \xi_{u}\\
  \xi_{d}\\
 \tilde a
\end{array}
\right), \quad
{\cal M}^{2} =
B\m\left(
\begin{array}{ccc}
\cot \b  & 1  &  -r\cos\b  \\
1  & \tan\b  &  -r\sin\b \\
  -r\cos\b  & -r \sin\b  &   r^{2}\cos\b \sin\b
\end{array}
\right),
\end{eqnarray}
where $r$ is defined by $r = v/f_{R}$.%
\footnote{In this study, we fix $B\m>0$ which is realized by using the definitions of R-symmetry and the Peccei-Quinn symmetry.}

As we expected, the above mass matrix possesses two massless states, i.e. rank$({\cal M}^{2})=1$.
One of them corresponds to the would-be Goldstone boson absorbed by the $U(1)_{Y}$
gauge boson, and the other one is the R-axion.
The mass eigenstates of the mass matrix are given by,
\begin{eqnarray}
\label{eq:massE}
&\left(
\begin{array}{c}
 G_{0}\\
  A_{0}\\
 a
\end{array}
\right)
=&
\left(
\begin{array}{ccc}
\sin \b  & -\cos\b  & 0 \\
\k\cos\b & \k\sin\b  &  -\k r\sin \b\cos\b \\
  \k r\cos^{2}\b\sin \b  & \k r \sin^{2}\b\cos\b  &   \k
\end{array}
\right)
\left(
\begin{array}{c}
 \xi_{u}\\
  \xi_{d}\\
 \tilde a
\end{array}
\right)
,\\
&(m_{G_{0}}^{2},m_{A_{0}}^{2},m_{a}^{2})& =\left(0,\frac{2B\m}{\k^{2}\sin2\b},0\right),
\end{eqnarray}
where $\k$ is defined by $\k = (1+ r^{2} \sin^{2} 2\b )^{-1/2}$.
As a result, the axial parts of the Higgs fields mix with the low energy R-axion;
\begin{eqnarray}
\label{eq:mixing}
\xi_{u} \sim \k r \cos^{2}\b\sin\b \times a, \quad\xi_{d} \sim \k r \sin^{2}\b\cos\b \times a.
\end{eqnarray}

The interactions of the R-axion with the SM fermions are the same found in the DFS axion
model\,\cite{Dine:1981rt}.
From the mixing in Eq.\,(\ref{eq:mixing}),  the coupling constants between the R-axion and
the SM fermions are given by,
\begin{eqnarray}
\label{eq:Rff}
\lambda_{u} = i y_{u}/\sqrt{2}\, r \cos^{2}\b \sin\b = i m_{u}/f_{R} \,\cos^{2}\b , \quad
\lambda_{d}= i y_{d}/\sqrt{2} \,r \sin^{2}\b \cos\b = i m_{d}/f_{R} \,\sin^{2} \b,
\end{eqnarray}
for up- and down-type quarks (and for their higher generation counterparts), respectively.
The leptons also couple to the R-axion in a similar way of the down-type quarks.
Unlike the case of the CP-odd Higgs scalar $A_{0}$, there
is no $\tan \b$ enhancement in coupling constants of the down-type quarks.

Notice that, for a given R-charge assignment to the $\m$-term,
the R-charges of the Higgs bosons are determined unambiguously so that
the would-be-goldstone boson $G_{0}$ is invariant under the R-symmetry.
That is, under the $U(1)_{R}$ symmetry, the axial fields $\xi_{u,d}$ are sifted by,
\begin{eqnarray}
 {\x_{u}}/v_{u} \to {\xi_{u}}'/v_{u} = {\xi_{u}}/v_{u} + X_{u}c, \quad
 {\x_{d}}/v_{d}\to {\xi_{d}}'/v_{d} = {\xi_{d}}/v_{d} + X_{d}c.
\end{eqnarray}
Then, by requiring that the would-be goldstone boson $G_{0}$ is invariant under the R-symmetry,
we obtain a condition,
\begin{eqnarray*}
 X_{u} \sin^{2}\b - X_{d} \cos^{2}\b = 0.
\end{eqnarray*}
By remembering the conditions in Eq.\,(\ref{eq:charge}), we find that the R-charges are given by,
\begin{eqnarray}
\label{eq:higgsR}
 X_{u} = 2 \cos^{2} \b, \quad  X_{d} = 2 \sin^{2} \b.
\end{eqnarray}
We can also check that $A_{0}$ is also invariant under the R-symmetry with the above charge assignment,
and only the R-axion $a$ in the low energy theory is shifted by,
\begin{eqnarray}
  {a} \to a' = a + 2 \k^{-1} f_{R} c.
\end{eqnarray}
This means that the decay constant of the R-axion below the electroweak scale is changed
from $f_{R}$ to
\begin{eqnarray}
f_{A}  = \k^{-1} f_{R} > f_{R},
\end{eqnarray}
although $\k \simeq 1$ and $f_{A} \simeq f_{R}$ for  $f_{R} \gg v$.

Finally, let us comment on the coefficient of the anomaly coupling in the low energy effective theory.
As we mentioned above, the coefficient $C_{H}$  depends on the model of the messenger sectors.
For example, in the case of  minimal gauge mediation with $N_{m}$ messenger pairs,
the coefficient is given by $C_{H} = -N_{m}$, while $C_{H} = 0$ in the
case of the second model discussed in the previous section.%
\footnote{The models proposed in Refs.\,\cite{Randall:1996zi,Seiberg:2008qj,Ibe:2007wp}
also give a vanishing coefficient, $C_{H}$.}
At  lower energy,  the coefficient gets additional contributions from heavy fermions such as gauginos and top quarks corresponding to the R-charges of their species.
Thus, at the scale below the gaugino masses and the top-quark mass, the coefficient of the anomaly term is changed to
\begin{eqnarray}
\label{eq:Ranom}
{\cal L} _{\rm eff} =
C_{L}\frac{g^{2}}{32\pi^{2}} \frac{a}{f_{A}}F \tilde F =
 (C_{H}+C_{2}(G) - \cos^{2}\beta)\frac{g^{2}}{32\pi^{2}} \frac{a}{f_{A}}F \tilde F,
\end{eqnarray}
which reproduces the anomaly of the R-symmetry by the shift of $a$.
Here $C_{2}(G)$ denotes the quadratic Casimir invariant which corresponds to the gauginos,
and we have used the charges of the Higgs bosons fixed in Eq.\,(\ref{eq:higgsR})
to determine the contribution from the top quarks in the third term.

At a lower energy scale, the lighter fermions also contribute to the anomaly coupling
once they are integrated out.
For example, for the R-axion with a hundred-MeV mass,
the value of the anomaly coefficient to photons below all hadron thresholds,  is given by,
\begin{eqnarray}
 C_{L,\gamma} = C_{H,1}+C_{H,2} + 2 - 9\left(\frac{4}{9}X_{u} + \frac{1}{9}X_{d}\right) - X_{d}
 -12\,{\rm tr}\left( Q_{A} Q_{\rm em}^{2}\right),
\end{eqnarray}
where the first two terms denote the messenger contributions
to the $U(1)_{Y}$ and $SU(2)_{L}$ gauge bosons,
the third term comes from the $SU(2)_{L}$ gaugino contribution, and the fourth from the tau lepton
contribution.
The final term comes from the light quark rotation with a charge assignment $Q_{A}$
which eliminates the anomalous coupling to the gluons\,\cite{Georgi:1986df},
and the explicit charge assignment of this manipulation is given by,
\begin{eqnarray}
 Q_{A} = \frac{C_{L, \rm gluon}}{2} M_{q}^{-1}/{\rm tr}(M_{q}^{-1})
 \simeq 0.16\,C_{L,\rm gluon},
\end{eqnarray}
where $M_{q}$ denotes the quark mass matrix
and we have used the mass ratio, $m_{u}/m_{d}\simeq 0.56$ in the last expression.
The coefficient $C_{L,\rm gluon}$ denotes
the anomalous coupling to the gluons at the low energy scale,
\begin{eqnarray}
C_{L,\rm gluon} = C_{H} + 3 -\frac{3}{2}\left(X_{u}+X_{d}\right).
\end{eqnarray}
Therefore, we obtain those coefficients,
\begin{eqnarray}
\label{eq:cagam}
C_{L,\gamma }&\simeq& C_{H,1} + C_{H,2} -2 - 1.9\, C_{H,3}, \cr
C_{L,\rm gluon} &\simeq& C_{H,3},
\end{eqnarray}
for a large value of $\tan\b$.

%%%%%%%%%%%%%%%%%%%%%%%%%%%%%%%%%%%%%%%%%%
\subsection*{Decay properties of R-axion}
In this subsection, we analyze the decay properties of the R-axion with a hundred-MeV mass
based on the effective Lagrangian obtained above.
In this mass region, the available decay modes of the R-axion are
those into a muon pair, an electron pair, a photon pair, and a gravitino pair.
As we will show below, the dominant decay modes are the one into the muons for
$m_{a}>2m_{\m}$, while the one into an electron pair dominates for $m_{a}<2m_{\m}$.

The R-axion decays into a pair of the SM fermions via the Yukawa interaction given in Eq.\,(\ref{eq:Rff})
and the decay width is given by,
\begin{eqnarray}
\G_{ff} = \frac{\l_{f}^{2}}{8\pi} m_{a}
\left(
1-\frac{4m_{f}^{2}}{m_{a}^{2}}
\right)^{1/2}
\,.
\end{eqnarray}
Thus, the decay rate and the decay length at the rest frame of the electron pair mode are given by
\begin{eqnarray}
\G_{ee} &\simeq& 3.1 \times 10^{-17}\,{\rm GeV}
\times \sin^{4}\b
\left( \frac{m_{a}}{300\,\rm MeV} \right)
\left( \frac{10^{4}\,\rm GeV}{f_{R}} \right)^{2}
\left(
1-\frac{4m_{e}^{2}}{m_{a}^{2}}
\right)^{1/2}
\,,\cr
c\tau_{ee} &\simeq& 6.3 \times 10^{2}\,{\rm cm}
\times \frac{1}{\sin^{4}\b}
\left( \frac{300\,\rm MeV} {m_{a}}\right)
\left( \frac{f_{R}} {10^{4}\,\rm GeV}\right)^{2}
\left(
1-\frac{4m_{e}^{2}}{m_{a}^{2}}
\right)^{-1/2}\,,
\end{eqnarray}
while the ones into a pair of the muons are given by,
\begin{eqnarray}
\label{eq:Rmumu}
\G_{\m\m} &\simeq& 1.3 \times 10^{-12}\,{\rm GeV}\times \sin^{4}\b
\left( \frac{m_{a}}{300\,\rm MeV} \right)
\left( \frac{10^{4}\,\rm GeV}{f_{R}} \right)^{2}
\left(
1-\frac{4m_{\m}^{2}}{m_{a}^{2}}
\right)^{1/2}\,,\cr
c\tau_{\m\m} &\simeq& 1.5 \times 10^{-2}\,{\rm cm}
\times \frac{1}{\sin^{4}\b}
\left( \frac{300\,\rm MeV} {m_{a}}\right)
\left( \frac{f_{R}} {10^{4}\,\rm GeV}\right)^{2}
\left(
1-\frac{4m_{\m}^{2}}{m_{a}^{2}}
\right)^{-1/2}\,.
\end{eqnarray}

The R-axion also decays into a pair of photons via the anomaly coupling as appeared in Eq.\,(\ref{eq:Ranom}).
The width of this mode is given by,
\begin{eqnarray}
\G_{\gamma\gamma} &\simeq& \frac{C_{L,\gamma}^{2}}{16\pi} \left(\frac{\alpha}{4\pi }\right)^{2}
 \left(\frac{m_{a}}{f_{R}}\right)^{2}m_{a}\,, \cr
 &\simeq&1.8\times 10^{-18}\,{\rm GeV}\times C_{L,\gamma}^{2}
\left(\frac{m_{a}}{300\,\rm MeV} \right)^{3}
\left(\frac{10^{4}\,\rm GeV}{f_{R}} \right)^{2}\,.
\end{eqnarray}
where, $C_{L,\gamma}$ is given in Eq.\,(\ref{eq:cagam}).
Compared with the decay widths into a pair of the electrons and muons  given above,
this process is subdominant for $m_{a}=O(100)\,$MeV.

Since we are considering the light gravitino scenario, the R-axion also decays into a gravitino pair,
$\psi\psi$, with the decay rate\,\cite{Bagger:1994hh},%
\footnote{For example, the helicity unsuppressed operators similar to the gaugino coupling
in Eq.\,(\ref{eq:RaxionL}) is proportional to $e^{-i a/f_{R}}\psi\psi$
at the leading order, which is forbidden in the massless gravitino and the exact R-symmetry limit.
}
\begin{eqnarray}
\G_{\psi\psi} &\simeq& \frac{1}{8\pi} \left(\frac{m_{3/2}}{f_{R}}\right)^{2} m_{a}\,,\cr
&\simeq& 1.2\times 10^{-26}\,{\rm GeV}
\left(\frac{m_{3/2}}{10\,\rm eV} \right)^{2}
\left(\frac{m_{a}}{300\,\rm MeV} \right)
\left(\frac{10^{4}\,\rm GeV}{f_{R}} \right)^{2}\,.
\end{eqnarray}
As a result, the decay mode into a pair of the gravitinos is also subdominant compared with
the modes into electrons and muons, as long as we are considering the light gravitino scenario,
$m_{3/2}=O(10)$\,eV.

Before closing this section, we comment on the decay properties of the R-axion with much heavier mass.
When the R-axion mass gets heavier than $3 m_{\pi}$, the decay mode into three pions is open.
Since the direct coupling between the R-axion and the $u, d$-quarks are suppressed
(see Eq.\,(\ref{eq:Rff})), the dominant contribution to the decay width of the three pion mode
comes from the anomaly coupling; $a F\tilde F$.
However, we found that the decay width into three pions is always suppressed compared
with the one into a muon pair for $m_{a}\lesssim 1$\,GeV by using an effective Lagrangian
of the R-axion$-3\pi$ interaction given in Ref.\,\cite{Chang:1993gm}.

For much heavier R-axion, the decay modes into heavier fermions,
such as the $\tau$ lepton and the bottom quark become available.
In such a region, the decay length is much shorter than the one discussed above.
Thus, the R-axion detection in the heavier mass region is more challenging at the LHC
than the case of the lighter R-axion with a mass in the hundred MeV range.

%%%%%%%%%%%%%%%%%%%%%%%%%%%%%%%%%%%%%%%%%%%%
\section{R-axion production at the LHC}\label{sec:production}
In this section, we consider R-axion production with a large transverse momentum at the LHC
via the gluon fusion processes $gg\to ga$, $qg\to qa$, and $q\bar{q}\to ga$.
These processes are similar to the CP-odd Higgs boson production processes in
the MSSM\,\cite{Kunszt:1991qe,Spira:1993bb}.
In those processes, the R-axion couples to the gluons via
the anomaly coupling given in Eq.\,(\ref{eq:RaxionL}) as well as
the triangle and box diagrams in which the gluino and the quarks circulate.
When the transverse momentum $p_{T}$ of the R-axion is smaller than the gluino mass
the anomaly couplings in the effective Lagrangian in Eq.\,(\ref{eq:Ranom})
give a good approximation to the gluino-loop diagrams.
Notice that the top-loop contribution is proportional to  $\cos^{2}\b$,
and hence, the top-quark contribution is suppressed for $\tan\b\gtrsim 2$
to the gluon fusion processes.
As for the contributions from the other lighter quarks,
they are also suppressed compared with the gluino contribution
when $p_{T}$ is much larger than the masses of those lighter quarks\,\cite{Baur:1989cm}.

From the above discussion,
we consider the R-axion production via the effective anomaly coupling,
\begin{eqnarray}
 {\cal L}_{\rm eff} =
C_{L}\frac{g_{3}^{2}}{32\pi^{2} } \frac{a}{f_{R}} F_{\m\n}\tilde{F}^{\m\n}=
  (C_{H}+ N_{c})\frac{g_{3}^{2}}{32\pi^{2} } \frac{a}{f_{R}} F_{\m\n}\tilde{F}^{\m\n},
\end{eqnarray}
where $N_{c}=3$ is a color factor.
This should give a good approximation for $p_{T}<m_{\rm gluino}$.
The matrix elements of the processes, $gg\to ga$, $qg\to qa$, and $q\bar{q}\to ga$,
from the above effective anomaly coupling are given in Refs.\,\cite{Kunszt:1991qe,Spira:1993bb}.

%%%%%%%%%%%%%%%%%%%%%%%%%%%%%%%%%%%%%%%%%%%%%
\begin{figure}[t]
 \begin{minipage}{.47\linewidth}
  \includegraphics[width=\linewidth]{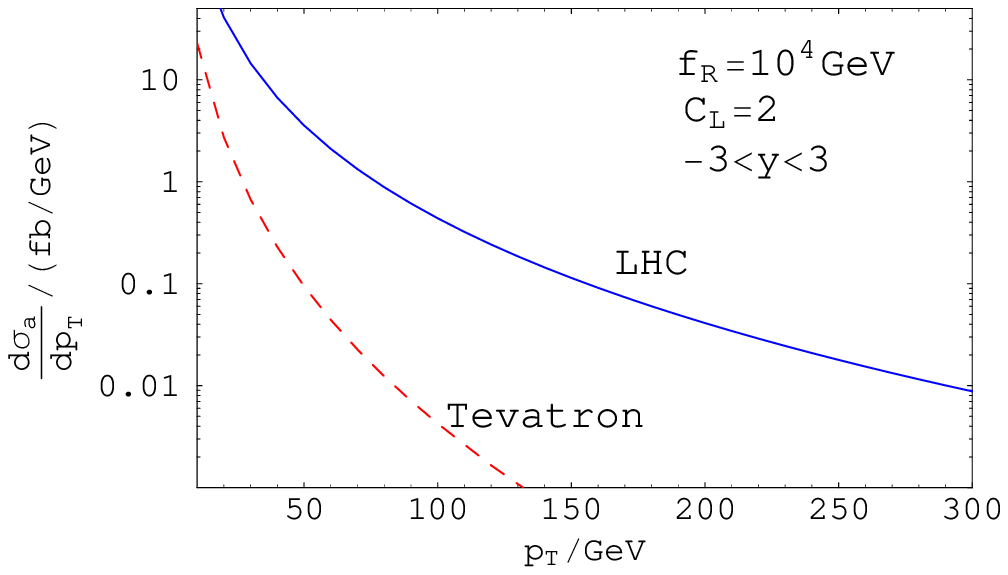}
 \end{minipage}
 \begin{minipage}{.515\linewidth}
  \includegraphics[width=1.0\linewidth]{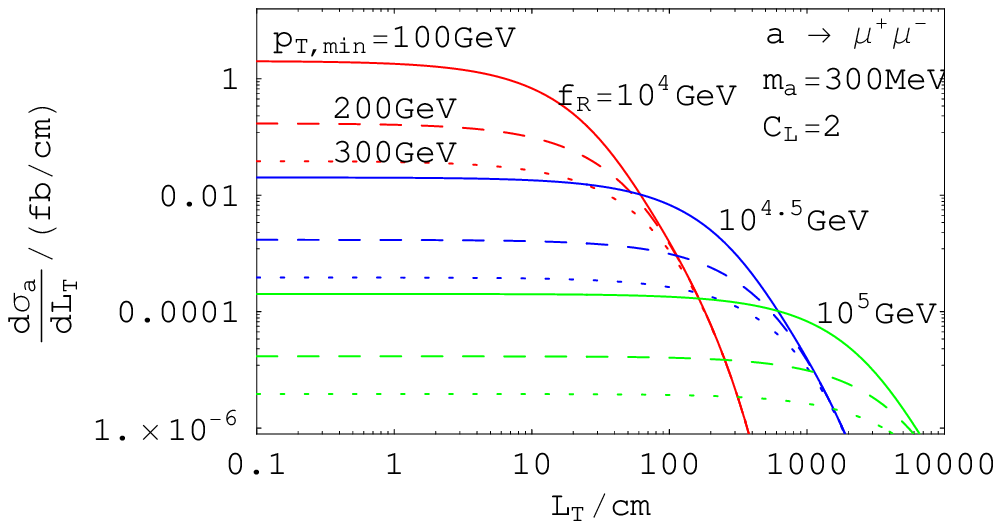}
 \end{minipage}
\caption{Left) Transverse momentum distribution of the R-axion with the mono-jet event
at the LHC (solid) and the Tevatron (dashed).
Here, we have taken $f_{R} = 10^{4}$\,GeV , $m_{a} = 300$\,MeV, and $C_{L}=2$.
We have integrated the rapidity $y$ for $|y|<3$.
Right)  Transverse decay length distribution of the R-axion decaying into a pair of the muons at the LHC.
The solid lines denote the distributions for the minimum transverse momentum, $p_{T,min} = 100$\,GeV
for $f_{R} = 10^{4,4.5,5}$\,GeV from the bottom up.
The dashed lines correspond to  $p_{T,min} = 200$\,GeV, and the dotted line to  $p_{T,min} = 300$\,GeV
for $f_{R} = 10^{4,4.5,5}$\,GeV from the bottom up.
}
\label{fig:production}
\end{figure}
%%%%%%%%%%%%%%%%%%%%%%%%%%%%%%%%%%%%%%%%%%%%%

In Fig.\,\ref{fig:production},
we present the transverse momentum distribution of the R-axion event with the mono-jet,
$d\s_{R}/dp_{T}$ at the LHC and the Tevatron, for $f_{R} = 10^{4}$\,GeV, and $m_{a} = 300$\,MeV.
In our analysis, we have used the CTEQ5 parton distribution functions\,\cite{Lai:1999wy}
with the QCD scale $\Lambda_{5}^{LO}=165.2$\,MeV.
The factorization and renormalization scales have been set to $p_{T}$.
The model dependent parameter $C_{L}$ has been set to $C_{L}=2$ which corresponds
to the minimal gauge mediation model with $N_{m}=1$ and $N_{m}=5$ (up to the sign of $C_{L}$).
The transverse momentum distribution is almost independent of the R-axion mass
as long as $m_{a}\lesssim 1$\,GeV, while it scales with $f_{R}^{-2}$ and $C_{L}^{2}$.
The figure shows that a sizable number of the R-axion events at the LHC even for $p_{T}\gtrsim 100\,$GeV,
while the cross section is rather suppressed at the Tevatron.

Now the question is how to detect the R-axion events.
As we have seen in the previous section, the R-axion has a long lifetime.
A useful variable is transverse decay length $L_{T} = \gamma c\tau \sin\theta$,
where $\gamma$ and $\theta$ are the boost factor and  the polar angle of the R-axion, respectively.
To see how the R-axion behaves after its production, we have plotted the transverse decay
length distribution of the R-axion decaying into a pair of muons for $m_{a}=300$\,MeV
in Fig.\,\ref{fig:production}.
The figure shows that the produced R-axion mainly decays at around $O(1)$\,cm,
$O(10)$\,cm, and $O(100)$\,cm, for $f_{R}=10^{4}$\,GeV, $f_{R}=10^{4.5}$\,GeV,
and $f_{R}=10^{5}$\,GeV, respectively.
Therefore, the R-axion leaves a displaced vertex from which a pair of the highly boosted muons is produced
inside the beam-pipe in the first case, the inner detector in the second case,
and the calorimeters in the last case.

\subsection*{R-axion reconstruction}
We now consider the reconstruction of the R-axion.
Since the two muons leave charged tracks inside the detector, it is possible to reconstruct
the displaced vertex of the R-axion decay as well as the R-axion momentum.
Notice that
 since the R-axion is highly boosted, the muons coming from the R-axion are collimated.
The typical azimuthal angle between the muons and the R-axion is given by,
\begin{eqnarray}
\label{eq:phiamu}
\phi_{a\m} \simeq \frac{m_{a}}{p_{T,a}},
\end{eqnarray}
where $p_{T,a}$ is the transverse momentum of the R-axion.
For $m_{a}=O(100)$\,MeV and $p_{T}=O(100)$\,GeV, the angle $\phi_{a\m}$ is expected to be $O(1)$\,mrad.
Therefore, the most important challenge to look for the R-axion is to
separate the two muon tracks with very small open angle.
Since a typical distance between two tracks at the pixel detectors is
of $O(100)\,\m$m, they can be separated by using the pixel detectors
in principle.
The actual separation efficiency and track resolution depend on the
detail of the trigger system of detectors.
In this study, we do not discuss the details of this problem, and
instead, we simply use the track resolutions of a single muon track
for each muon.
At the ATLAS detector, for example, the resolution of the azimuthal angle of the muon is approximately
given by,
\begin{eqnarray}
\label{eq:azimuth}
\sigma_{\phi} \simeq 0.075\oplus \frac{1.8}{p_{T}/{\rm GeV} \sqrt{\sin\theta}}\,\, ({\rm mrad}),
\end{eqnarray}
where $\theta$ is the polar angle of the R-axion.
Thus, the azimuthal directions of the two muons can be well measured\,\cite{:1999fq}, and
they are well separable.

Related to the angle between the muons and the R-axion,
the transverse impact parameters $d_{0}$ of each muons are also expected to be
 \begin{eqnarray}
 \label{eq:impact}
 d_{0} \simeq \,\frac{p_{T,a}}{m_{a}}\, c\tau_{\m\m} \,\phi_{a\m}
 \simeq \, c\tau_{\m\m}.
\end{eqnarray}
The typical decay length at the rest frame of the R-axion given in Eq.\,(\ref{eq:Rmumu})
is of the order of  $100\,\m$m over a large parameter range.
Therefore, the finite impact parameters of the two muons can be measured at the ATLAS
detector where the resolution of the transverse impact parameter is given by\,\cite{:1999fq},
\begin{eqnarray}
\label{eq:resolution}
\sigma_{d_{0}} \simeq 11 \oplus \frac{73}{p_{T}/{\rm GeV} \sqrt{\sin\theta}} \quad(\m{\rm m}).
\end{eqnarray}

Once the transverse impact parameter is measured, the transverse decay length is then determined by,
\begin{eqnarray}
 L_{T} = \frac{d_{0}}{\sin( \phi_{a\m})}.
\end{eqnarray}
This is expected to be several cm as shown in Fig.\ref{fig:production}.
From the resolutions of the impact parameter and the azimuthal angle, the resolution
of the transverse decay length is estimated by,
\begin{eqnarray}
\frac{\sigma_{L_{T}}}{L_{T}} = \frac{\s_{d_{0}}}{d_{0}}
 \oplus \frac{\s_{\phi}}{\phi_{a\m}}.
\end{eqnarray}
Therefore, for $d_{0}=O(100)\,\m$m and $\phi_{a\m}=O(1)\,$mrad,
we expect the resolution of the transverse decay length
to be marginal to better than 10$\,\%$.

Before closing this section, we mention the R-axion mass measurement.
In terms of the observed momentum, the mass of the R-axion is given by,
\begin{eqnarray}
 m_{a}^{2} \simeq 4 m_{\m}^{2} + p_{\m} p_{\bar{\m} }\, \theta_{\m\m}^{2}.
\end{eqnarray}
where $p_{\m(\bar{\m})}$  denote the size of the momentum
and the transverse momentum of the muon (anti-muon).
The absolute angle between two muons , $\theta_{\m\m}$,
is given by,
\begin{eqnarray}
\label{eq:thmumu}
\theta_{\m\m}  \simeq \sqrt{  {\mit \Delta}\theta^{2}   + \sin^{2}\bar\theta\, {\mit \Delta}\phi^{2}},
\end{eqnarray}
where ${\mit \Delta}\theta$ and ${\mit \Delta}\phi$ denote the differences of the
polar and the azimuthal angles of the muon and anti-muon, while $\bar\theta$
is the average of the muon polar angles.
Since the R-axion is highly boosted, the expected value of  $\theta_{\m\m}$ is given by,
\begin{eqnarray}
\label{eq:thetamumu}
 \theta_{\m\m} \simeq 2 \sin\bar\theta\,\, \frac{m_{a}}{p_{T,a}}.
\end{eqnarray}
At the ATLAS detector, by using the resolution of the  azimuthal angle in Eq.\,(\ref{eq:azimuth})
and the one of the polar angle\,\cite{:1999fq},
\begin{eqnarray}
 \sigma_{\cot\theta}  \simeq 0.70\times 10^{-3}\oplus \frac{2.0\times 10^{-3}}{p_{T}/{\rm GeV} \sqrt{\sin\theta}},
\end{eqnarray}
the resolution of the angle between muons is given by,
\begin{eqnarray}
\frac{\sigma_{\theta_{\m\m}}}{\theta_{\m\m}} &\simeq&
\sin^{2}\bar\theta\, \frac{{\mit \Delta} \theta}{\theta_{\m\m}^{2}}
\,\sigma_{\cot\theta} \,
\oplus \,
\sin^{2}\bar\theta\, \frac{{\mit \Delta} \phi}{\theta_{\m\m}^{2}} \,\sigma_{\phi}
\,\oplus\, (\m \to \bar \m),\cr
&\simeq&\sin\bar\theta \,\frac{p_{T,a}}{2m_{a}}\,\sigma_{\cot\theta} \,
\oplus \,\frac{p_{T,a}}{2m_{a}}\,\sigma_{\phi}
\,\oplus
\,(\m \to \bar \m),
\end{eqnarray}
where we have used Eq.\,(\ref{eq:thmumu}) and ${\mit \D}\theta = 2 m_{a}/p_{a}$,
and ${\mit \D}\phi = 2 m_{a}/p_{T,a}$.
Thus, the resolution of the absolute angle is expected to be of $O(10)$\,\%.
Therefore, the R-axion mass measurement is possible, although it requires careful study.

%%%%%%%%%%%%%%%%%%%%%%%%%%%%%%%%%%%%%%%
\section{Background Estimation}\label{sec:background}
In this section, we consider relevant background for the
R-axion$\,+\,$jets events. As we have discussed, we select events
with a jet of $p_{T}>100$\,GeV balanced by a low-mass muon pair
emerging from a displaced vertex. First of all, a potential
background can come from mis-measuring the impact parameters of
a prompt muon pair. The probability depends on the decay length of
the axion and the tracking precision. Since we require the
transverse impact parameters of both the muons to be around
$d_{0}$ in Eq.\,(\ref{eq:impact}), the mis-measurement probability
$P_{d_{0}}$ is roughly estimated by,
\begin{eqnarray}
\label{eq:pd0}
 P_{d_{0}} \simeq { \rm Erfc}^2\left(\frac{d_{0}}{\sqrt{2}\s_{d_{0} } } \right),
\end{eqnarray}
where $\s_{d_{0}}$ is the resolution of the transverse impact
parameter given in Eq.\,(\ref{eq:resolution}). Thus, the
probability $P_{d_{0}}$ is quite small for the typical transverse
impact parameter of the R-axion, $d_{0} = O(100)\,\m$m, i.e.
$P_{d_{0}}\ll 10^{-5}$. The limit of the suppression factor
corresponds to $d_{0}\sim 45 \m$m, about 3$\sigma$ from primary
interaction vertex.

The production cross section of a prompt muon pair with $p_T>50$ GeV
is generically less than $10$\,pb. For example, the light
unflavored mesons, such as $\rho, \omega, \phi, \eta$ can promply
decay to two muons. The cross section of these meson with
$p_T>100$ GeV is similar to that of two jet event which is roughly
$\sigma_{\rm meson}= 1\,\m$b and the two muon decay branching
ratio is $Br^{(\rm meson)}_{\m\m}=O(10^{-5})$ when taking into
account the muon isolation. The muon pair can also be produced via
the Drell-Yan process. The production cross section of such events
is also about $O(10)$\,pb for $p_{T}\gtrsim 100$\,GeV. Hence, the
cross section of the background event is less than about $0.1$ fb
which is much smaller than that of the signal (see
Fig.\ref{fig:detectability} at the end of this section).

More serious background comes from events which have a non-prompt
muon pair. The main candidates for the background are as follows;
\begin{itemize}
    \item A muon pair in heavy flavor ($B$, $D$) meson decays.
    \item A muon pair in light flavor ($K$) meson decays.
    \item A muon pair in gluon splitting followed by meson decay.
    \item A photon conversion in the detector material to a muon pair.
\end{itemize}
Among these, the most serious background comes from the muon pair
in a cascade decay of the $B$ meson. The muon pairs from a cascade
decay of the $K$ meson and photon conversions will also be
important if we look for the displaced vertex outside of the
beam-pipe. We will estimate these background in turn in the
following subsection.

%%%%%%%%%%%%%%%%%%%%%%%%%%%%%%%%%%%%%%%%%%%%%%%
\subsection{A muon pair from heavy meson decay}\label{sec:Bmeson}
The muon pairs from the heavy meson decays are serious background,
since the decay length of these mesons are similar to that of the
R-axion, i.e. $c\tau =O(100)\,\m$m. Among them, some serious
contributions are as follows;
 \renewcommand{\labelitemi}{$\ast$}
\begin{itemize}
    \item A muon pair in non-resonant decays.
    \item A muon pair in cascade decays.
    \item A fake muon contribution in semi-leptonic meson decay.
\end{itemize}

%%%%%%%%%%%%%%%%%%%%%%%%%%%%%%%%%%%%%%
\begin{table}
  \centering
\begin{tabular}{|c|c|c|c|}
  \hline
  process &  $Br^{(X)}_{\m\m}$ &  $P_{\rm geo}$ & $\sigma_{ X \to \mu\mu}$(fb) \\
  \hline
  $B^{0}\to K^{*0}\m^{+}\m^{-}$ & $1.3\times 10^{-6}$\,\cite{Melikhov:1997wp}
 &  $(m_a/m_B)^3$ & $10^{-4}$  \\
  $B^{0}\to J/\psi + X \to \m^{+}\m^{-}+X$ &  $\simeq 5.9\times 10^{-5}$ & $ \lesssim 10^{-6}$ & $10^{-5}$ \\
  $B^{0}\to D^{0}+X \to D^{0}+\m^{+}\m^{-}$  &  $< 10^{-8} $ & $(m_a/m_B)$ & $10^{-4}$ \\
  $B^{0}\to D^{\pm}+\mu^{\mp} +\nu \to \m^{+}\m^{-}+X$  & $10^{-2}$ & $(m_a/m_B)^3$ & 1 \\
  $B^{0}\to\pi^{-}\m^{+}\n$ &  $3 \times 10^{-8}$ &  $(m_a/m_B)^3$ & $10^{-5}$ \\
     \hline
  $D^{0}\to \rho^{0}+\m^{+}\m^{-}$  & $1.5 \times 10^{-7}$ \cite{Buchalla:2008jp} &  $(m_{a}/m_{D})^{3}$ & $10^{-3}$ \\
  $D^{0}\to \omega+K_{S}^{0}\to \m\m+ K_{S}^{0}$ & $10^{-6}$ & $(m_{a}/m_{D})$ & $1$ \\
  $D^{0}\to \rho^{0}+\pi^{0}\to \m\m+\pi^{0}$ &$10^{-7}$ & $(m_{a}/m_{D})$ & $10^{-1}$ \\
  $D^{0}\to K^{\pm}+\m^{\mp}+\n$ &  $10^{-5}$  & $(m_{a}/m_{D})^{3}$ & $10^{-1}$ \\
  $D^{0}\to \pi^{\pm}+\m^{\mp}+\n$  & $6\times10^{-7}$ &  $(m_{a}/m_{D})^{3}$ & $10^{-2}$ \\
  \hline
\end{tabular}
\caption{Summary of rough estimation of muon pair background from
different heavy meson decay modes. Here we have use
$m_X=(5,2)$\,GeV and $c\tau=(460,123)$\,$\m$m for B and D.
As our signal cross section is about $10$\,fb, the
background will be claimed insignificant if it's cross section is
less than $0.1$\,fb.
The final cross sections from the $B^{0}\to D^{\pm}+\mu^{\mp} +\nu \to \m^{+}\m^{-}+X$
and  $D^{0}\to \omega+K_{S}^{0}\to \m\m+ K_{S}^{0}$ modes
are further suppressed than given here (see discussion in the text).
}
\label{bgtable}
\end{table}
%%%%%%%%%%%%%%%%%%%%%%%%%%%%%%%%%%%%%%

Non-resonant decays are inclusive meson decays of the type
$(B,D)\to \m^{+}\m^{-}+X$. The Standard Model prediction of the
non-resonant inclusive $B$ meson decay into a muon pair is highly
suppressed. The most serious mode is, $B^{0}\to
K^{*0}\m^{+}\m^{-}$. For the cascade decays, $(B,D)$ first decays
to some resonances and followed by the decay of one of the
resonances. When the $\pi$ and $K$-meson decay in flight or punch
through to the muon chamber, they can fake the muon. Thus, we
should also consider semi-leptonic decay of heavy flavor meson
with one of the light hadron ($\pi$, $K$) faking the muon.

The effective cross section of each of these potential background
events from meson $X=(B,D)$ can be described by
\begin{eqnarray}
\label{eq:NR}
 \sigma_{ X \to \mu\mu} \,&<&\, \sigma_{\rm X}\times Br^{(X)}_{\m\m} \times P_{L_{T}}\times  P_{\rm
 geo}\times P_{\slashchar b}\times P_{\slashchar{jet}},\cr
    \,&<&\, Br^{(X)}_{\m\m} \times P_{L_{T}}\times  P_{\rm
 geo}\times P_{\slashchar b} \times {\rm nb},
\end{eqnarray}
where we have used the fact that $(b,c)+100\,$GeV jet cross
sections are about the same and they are
$\sigma_{B,D}\simeq10$\,{\rm nb} with $p_{T}\gtrsim
100$\,GeV\,\cite{Nason:1999ta}. A factor $P_{\slashchar{jet}}$
represents the muon isolation condition that require no
significant hadronic activity in the detector. It is generally
estimated to be around $10^{-1}$. A factor $Br^{(X)}_{\m\m}$ is
the effective branching ratio of $\m^+\m^-$ decay channel. As we
have to also take into account of the fake muon, this branching
ratio should also include semi-leptonic decay with one of the
light meson faking the muon. In the case of fake muon, the fake
rate $P_{\m/(\pi,K)}$ should be included in the effective
branching ratio
\begin{equation}
    Br^{(X)}_{\m\m}=Br^{(X)}_{(\pi,K)\m}\times P_{\m/(\pi,K)}.
\end{equation}
A rough estimation of the fake muon probability $P_{\m/(\pi,K)}$
is given by,
\begin{eqnarray}
P_{\m/X} = P_{\rm mis-id}\times Br_{X\to\m+\n}\times \int^{r_{\rm
out}}_{0} dL_{T} \frac{1}{c\tau_{X}} \frac{m_{X}}{p_{T,X}}
\exp\left[ - \frac{m_{X}}{p_{T,X}} \frac{L_{T}}{c\tau_{X}}\right]
\times
n_{X}(L_{T}),
\quad (X = \pi^{\pm}, K^{\pm}),
\end{eqnarray}
where $P_{\rm mis-id}$ is the probability of the misidentification
of the mesons to the muon, $n_{X}(L_{T})$ is the rate of the
punch-through mesons reaching to the transverse length $L_{T}$,
and the $r_{\rm out}$ is the outer radius of the calorimeter. In
the case of the ATLAS detector, the calorimeter extends from an
inner radius about $2$\,m to and outer radius $4$\,m, and the
total thickness of the calorimeter is about 11 interaction
lengths. By using these parameters, the punch-through rate at a
given length $L_{T}$ is given by,
\begin{eqnarray}
 n_{X}(L_{T}) \simeq 1- \theta(L_{T}-r_{\rm in})\int_{r_{\rm in}}^{L_{T}}dx \frac{11}{{\mit \D}r_{\rm calo}}
 \exp \left[ -11\frac{ (x-r_{\rm in})}{{\mit \D}r_{\rm calo}}
 \right], \quad (r_{\rm in}\simeq 2\,{\rm m},\,  r_{\rm out}\simeq 4\,{\rm m},
 \, {\mit \D}r_{\rm calo}=r_{\rm out}-r_{\rm in} ).
\end{eqnarray}
As a result, we obtain rough estimations of the muon fake rates,
\begin{eqnarray}
\label{eq:fake}
 P_{\m^{\pm}/\pi^{\pm}} \lesssim 2\times 10^{-4}, \cr
  P_{\m^{\pm}/K^{\pm}} \lesssim 4\times 10^{-4},
\end{eqnarray}
for $p_{T}\gtrsim 50$\,GeV. In the above expression, we have
assumed $P_{\rm mis-id} = 50$\,\% for $\pi^{\pm}$ and $P_{\rm
mis-id} = 10$\,\% for $K^{\pm}$\,\cite{Cataldi:2006cx}.

The suppression factor $P_{L_{T}}$ represents the difference of
the expected transverse decay length of the R-axion and the
mesons. It can be estimated as $P_{L_T}\sim L_T/L_X$ where $L_T$
and $L_X$ are the transverse decay lengths of R-axion and meson
$X$ respectively. The factor $P_{\rm geo}$ parameterizes a
rejection factor based on the difference of the event geometry of
the $X$ meson decay and the R-axion decay. The factor $P_{\rm
geo}$ is estimated in the following way. A typical absolute angle
between the muons from $X$ meson decay is roughly given by,
\begin{eqnarray}
\theta_{\m\m}^{X} \simeq 2 \sin\bar\theta\, \frac{m_{X}}{p_{T,X}},
\end{eqnarray}
where $p_{T,X}$ is the transverse momentum of the $X$ meson and
$m_{X}$ is the mass of the meson. Thus, by comparing the typical
angle between the two muon from R-axion decay in
Eq.\,(\ref{eq:thetamumu}), the geometrical suppression factor is
roughly given by,
\begin{eqnarray}
\label{eq:geo}
 P_{\rm geo} \simeq \left( \frac{\theta_{\m\m}}{\theta_{\m\m}^{X}}\right)^{3} \simeq
  \left( \frac{p_{X}}{p_{a}}   \frac{m_{a}}{m_{X}}\right)^{3}
  \simeq
  \left( \frac{p_{T,X}}{p_{T,a}}   \frac{m_{a}}{m_{X}}\right)^{3}
    \simeq
  \left(  \frac{m_{a}}{m_{X}}\right)^{3}.
\end{eqnarray}
In the final equality, we have also used $p_{T,X}\simeq p_{T,a}$,
which is required for the transverse momentum balance between the
muon pair and the remaining jets. In the above expression, the two
powers of $({\theta_{\m\m}}/{\theta_{\m\m}^{B}})$ out of three
come from the requirement that the absolute angle of two muons
should be as narrow as $\theta_{\m\m}$, while the remaining power
comes from the requirement that the transverse impact parameters
of muons for a given transverse decay length are $O(c\tau_{\m\m})$
given in Eq.\,(\ref{eq:Rmumu}).

A factor $P_{\slashchar b}$ is a rejection factor obtained from
vetoing a $b$-jet which does not fake the muon pair. For example,
the $b$-tagging efficiency can be about $90$\,\%, while the
misidentification rate of a gluon jet as a $b$-jet is about
$0.1$\,\cite{:1999fq}. Thus, we can suppress a background event by
a factor of $0.1$, while keeping the efficiency of the signal
event at about 90\,\%.

The basic analysis given above does not include the momentum
balance that requires the muon pair to acquire large portion of
the jet momentum in order to pass the $p_T$ cut and that the sum
of momentum of the muon pair points back to the primary vertex. We
will use this basic analysis to eliminate most of the background
and to identify dominant background which can not be eliminate by
this basic cut. Since our signal cross section is about 10 fb, we
will claim the background insignificant if its cross section is
less than $0.1$ fb. For those backgrounds which are lager than
$0.1$ fb, we will treat them more carefully in the next step. For
$m_{a} \lesssim 1$\,GeV, where the displaced vertex is
$L_{T}=1-5$\,cm, we
summarize the estimation of various backgrounds in Table\,\ref{bgtable}.%
\footnote{In the table, we have not included the exclusive decay
mode $B\to \m^{+}\m^{-}$ whose branching ratio in the Standard
Model is too small
($Br_{\m\m}^{(B)}=O(10^{-9})$\,\cite{Buras:2003td}) to give a
serious background. In the table, we have also used $P_{\rm
geo}\lesssim 10^{-6}$ for $m_{a}\lesssim 1$\,GeV in the $J/\psi$
mode. This comes from the fact that the exclusive decay $J/\psi\to
\mu\mu$ is a two body decay, and hence, the narrow opening angle
between the two muons in Eq.\,(\ref{eq:phiamu}) is difficult to
achieve. }

As been illustrated in Table.\,\ref{bgtable},  a muon pair coming
from a cascade decay of the $B$ meson, $B^{0}\to
D^{\pm}+\m^{\mp}+\n\to K^{0} + \m^{+}\m^{-} + \n\n$ is the
dominant background. Note that, although the subsequent $D$ meson
decay also leaves a displaced vertex, the typical distance of the
vertex from the muon track from the $B$-decay is too short to be
distinguished. For these background events, we need more careful
analysis in order to distinguish them form the signal. For the
cascade decay of $D$ meson, the background is about $1$ fb.
However, the muon-pair isolation cut in a $c$-jet is expected to
be more efficient than that in a $b$-jet, since the $D$ meson
momentum fraction to the $c$-jet is typically much smaller than
the one of the $B$ meson to the $b$-jet. Thus, although we do not
pursue further analysis, we expect that the background cross
section from $D^{0}\rightarrow w+K_S^{0}$ is more suppressed than
that shown in Table. (\ref{bgtable}). Further suppression can also
be achieved by more detailed analysis similar to that given below
for the $B$ decay.

%%%%%%%%%%%%%%%%%%%%%%%%%%%%%%%%%%%%%%%%%%%%%
\begin{figure}[t]
% \begin{minipage}{.45\linewidth}
%  \includegraphics[width=\linewidth]{rhoMET.eps}
% \end{minipage}
 \begin{minipage}{.45\linewidth}
  \includegraphics[width=\linewidth]{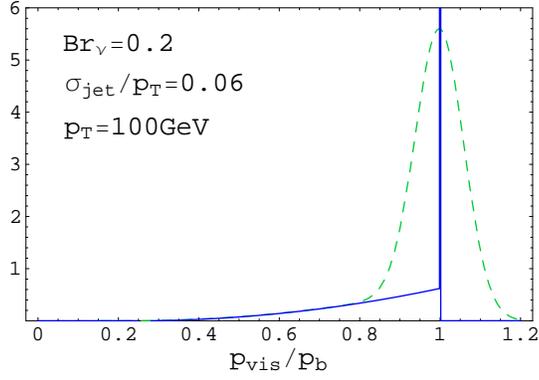}
 \end{minipage}
\caption{The momentum fraction of the visible energy of a $b$-jet.
The solid line denotes the visible momentum, i.e. $p_{b}-p_{\n}$
before smearing, while the dashed line includes the smearing
effect. In the figures, we have assumed the inclusive neutrino
mode of the $b$-jet is $20$\,\%, and the jet momentum resolution
about $6$\,\% for $p_{T}\simeq 100$\,GeV. } \label{fig:lightmeson}
\end{figure}
\begin{figure}[t]
 \begin{minipage}{.49\linewidth}
  \includegraphics[width=.93\linewidth]{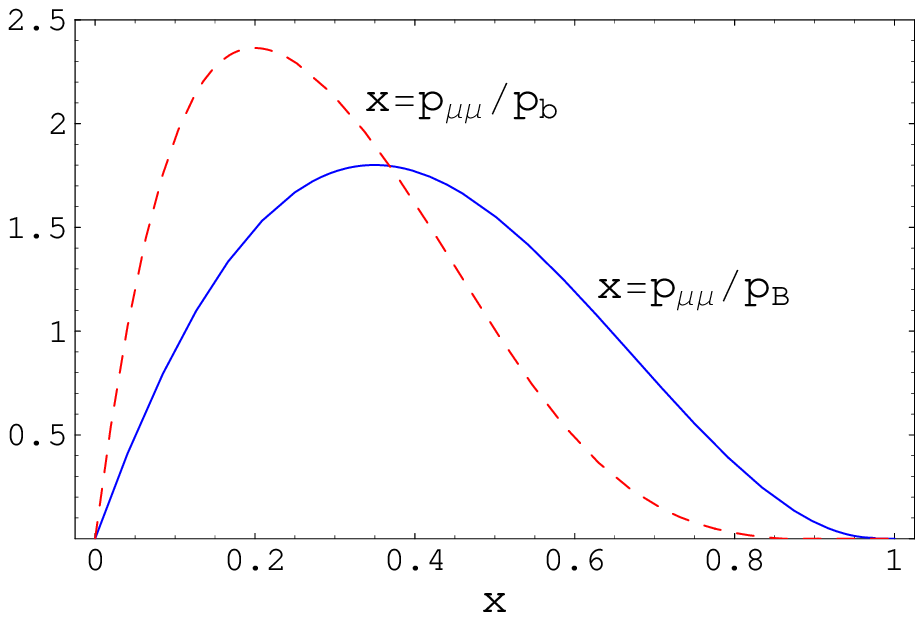}
 \end{minipage}
 \begin{minipage}{.49\linewidth}
  \includegraphics[width=\linewidth]{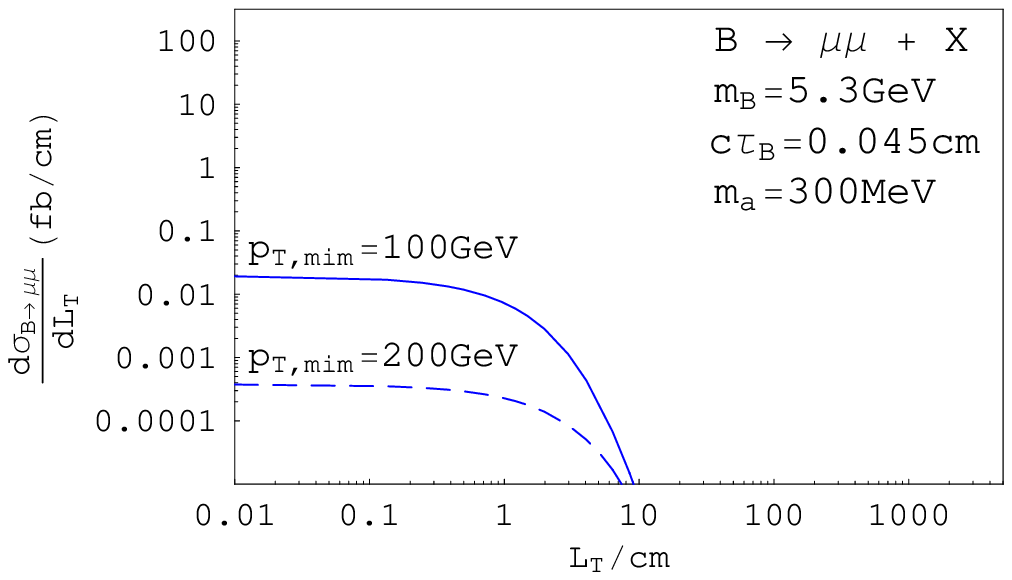}
 \end{minipage}
\caption{
Left) The momentum fraction of a muon pair to the $B$ meson momentum (solid line)
and to the $b$-quark momentum (dashed line)
in a cascade decay $B^{0}\to D^{\pm} \m^{\mp}\n \to K^{0} \m\m\n\n$.
In this analysis, we have assumed spherical decay of $B$- and $D$ mesons.
Right) Transverse decay length distribution of the $B$ meson decaying into a pair of muons
which fakes the muons from the R-axion decay.
The solid (dashed) line denotes the distribution with a lower cut on the transverse momentum
of the R-axion candidate, $p_{T,min}=100\,(200)$\,GeV.
The distribution scales by $(m_{a}/300\,{\rm MeV})^{3}$ for a different R-axion mass.
}
\label{fig:bjet}
\end{figure}

%%%%%%%%%%%%%%%%%%%%%%%%%%%%%%%%%%%%%%%%%%%%%
In addition to the suppression discussed above, the background
from $B$ meson cascade decay can be suppressed further by
requiring transverse momentum balance between the muon pair and
the remaining jets. In Fig.\,\ref{fig:bjet}, we show the momentum
fraction of the muon pair in the cascade decay, where we have
assumed spherical decays of the $B$ and $D$ mesons. To obtain the
$x=p_{\m\m}/p_{B}$ distribution, we have used the $p_{B}/p_{b}$
distribution given in\,\cite{Abbiendi:2002vt,Abe:2002iq}. The
figure shows that the spectrum of the muon pair with
$p_{\m\m}/p_{b}>0.7$ is highly suppressed. Also, the other jet
which does not involve the muon pair may decay into neutrinos. In
Fig.\,\ref{fig:lightmeson},  we have plotted the visible energy
fraction of the $b$-jet, $p_{\rm vis}=p_{b}-p_{\n}$. In the
figure, we have assumed the inclusive neutrino mode of the $B$
meson, $B\to D+\n+{\rm anything}$,  is $20$\,\%. From those
distributions, we can obtain a rough estimation of the size of the
missing transverse momentum. As a result, we found that the
background is  rejected by a factor better than $10^{-2}$, by
requiring that the missing transverse momentum is smaller than
$20$\,\% of the observed $p_{T}$ of the muon pair. On the other
hand, the signal efficiency by the missing transverse momentum
cut, $\slashchar p_{T}<0.2\, p_{T,\m\m}$, is close to $1$, since
the main cause of the missing transverse momentum in the signal
event is the resolution of the jets, $\sigma_{\rm jet}\sim
0.6\,{\rm GeV}\times\sqrt{p_{T}/{\rm GeV}}$\,\cite{:1999fq}. Thus,
for a conservative estimation, we can use $P_{\slashchar
E_{T}}\simeq 10^{-2}$.

Putting all these suppression factors together, the transverse
decay length distribution of the background events is roughly
estimated by,
\begin{eqnarray}
\label{eq:BtoD} \frac{ d\sigma_{ B \to \mu\mu }}{dL_{T}} \,<\,
Br^{(B)}_{\m\m}\times
 P_{\slashchar E_{T}} \times  P_{\slashchar b}
 \times
  \int_{p_{T,min}}\hspace{-.8cm} dp_{T}\,
    \frac{1}{ c\tau_{B} } \,
   \frac{m_{B}}{p_{T,B}  } \,
\exp\left[
 - \frac{m_{B}}{p_{T,B}}
  \frac{L_{T}}{ c\tau_{B}}
  \right]\,
    \left(   \frac{m_{a}}{m_{B}}\right)^{3}
  \,
  \frac{d\s_{B}}{dp_{T,B}}\,,
\end{eqnarray}
where $P_{\slashchar E_{T}}\simeq 10^{-2}$ and $P_{\slashchar
b}\simeq 10^{-1}$. In Fig.\ref{fig:bjet}, we show the resultant
transverse decay length distribution for $m_{a}=300$\,MeV. The
figure shows that the final suppression factor in
Eq.\,(\ref{eq:BtoD}) is about $10^{-3}$ for $L_{T}\simeq 1$\,cm.
In the analysis, we have used the production cross section of the
$b$-quark pair production given in Ref.\,\cite{Nason:1999ta} with
$p_{B}=0.8\,p_{b}$. By comparing with the transverse decay length
distribution of the R-axion in Fig.\ref{fig:production}, it is
found that the signal-to-background ratio is good especially for
$L_{T}>O(1)$\,cm and $m_{a}\lesssim 1$\,GeV.%
\footnote{ In this analysis, we have not used a muon pair
isolation cut on the background events. This requirement is not
independent of the momentum balance requirement. }
A simple numerical
simulation of the $P_{\mu^+\mu^-}$ distribution in b cascade decay
using CalcHEP\cite{Pukhov} is also carried out and used to
estimate the rejection factor. By imposing the cut that the
invariant mass of the two muons is less than 1 GeV and the $p_T$
of both muons is larger than 40 GeV, the total cross section of
$pp\rightarrow b\bar{b}\rightarrow \mu^+\mu^-$ is about $0.15$ fb
using $Br^{(B)}_{\m\m}=10^{-2}$. After including the factor
$P_{\slashchar b}\sim 0.1$, this background can be suppressed to
less than $0.015$ fb.

%%%%%%%%%%%%%%%%%%%%%%%%%%%%%%%%%%%%%%%%%%%%%
%%%%%%%%%%%%%%%%%%%%%%%%%%%%%%%%%%%%%%%%%%%%%
%%%%%%%%%%%%%

%%%%%%%%%%%%%%%%%%%%%%%%%%%%%%%%%%%%%%%
\subsection{A muon pair from $K$ meson decay}\label{sec:Kaon}
The $K$ mesons have long lifetimes: $c\tau_{K} = 15.3$\,m for
$K_{L}^{0}$ and $c\tau_{K} = 2.68$\,cm for $K_{S}^{0}$.
Their inclusive decays into a
pair of  (fake) muons contribute to the background, if they decay
at a short distance. The dominant decay modes which contribute the
above events are, $K_{L}^{0}\to \pi^{\pm}+\m^{\mp}+\n\,
(Br=27\,\%)$, $K_{S}^{0}\to \pi^{\pm}+\m^{\mp}+\n\, (Br=4.7\times
10^{-4})$, and $K_{S}^{0}\to \pi^{+}+\pi^{-}\, (Br=68\,\%)$. Since
those mode requires $\pi\to\m$ misidentification, the effective
branching ratio is given by,
\begin{eqnarray}
 Br^{\rm (eff)}(K_{L}^{0}\to \m(\pi)+\m+\n) &=&
  \left( \frac{m_{K} }  {p_{T,K}}
   \frac{ r_{\rm pipe}}  {c\tau_{K_{L}}}
  \right)
  \times P_{\m/\pi}\times
 Br(K_{L}^{0}\to \pi+\m+\n)\simeq 10^{-9}, \cr
 Br^{\rm (eff)}(K_{S}^{0}\to \m(\pi)+\m+\n) &=&
 \left( \frac{m_{K} }  {p_{T,K}}
   \frac{ r_{\rm pipe}}  {c\tau_{K_{S}}}
  \right)
    \times P_{\m/\pi}\times
 Br(K_{S}^{0}\to \pi+\m+\n)\simeq 10^{-9}, \cr
 Br^{\rm (eff)}(K_{S}^{0}\to \m(\pi)+\m(\pi)) &=&
  \left( \frac{m_{K} }  {p_{T,K}}
   \frac{ r_{\rm pipe}}  {c\tau_{K_{S}}}
  \right)
    \times P_{\m/\pi}^{2}\times
 Br(K_{S}^{0}\to \pi+\pi)\simeq 3\times 10^{-10},
\end{eqnarray}
for $p_{K,T}\gtrsim 100$\,GeV and $r_{\rm pipe}\simeq 5$\,cm. The
cross section of the background is also suppressed by the
requirement of the momentum balance and the muon pair isolation.
For example,  if we are require $p_{K}/p_{\rm jet}>0.7$, the
background is suppressed by $O(10^{-2})$ (see the $K$ meson
spectrum in a light jet given in Ref.\,\cite{Shlyapnikov:2001jf}).

Therefore, the background events are suppressed by
$10^{-(11-12)}$, and hence, the background from the $K$ meson
decay is subdominant for $p_{T}\gtrsim 100$\,GeV where the cross
section of jets is less than about $1\,\m$b. As a result, we find
that the background cross section from the $K$ meson decay is also
subdominant.

 %%%%%
\subsection{Heavy flavor meson pair from gluon splitting}
When the gluon splits into a pair of b quarks, we may have two $B$
mesons in a single jet. In this case, both $B$ and $\bar B$ can
decay to muons, and this muon pair can contribute to the
background if the two tracks intersect at the finite transverse
length. The suppression factors to the background cross section of
the accidental muon pair are as follows.
 \renewcommand{\labelitemi}{$\ast$}
\begin{itemize}
\item $Br_{\m+X}^{2}$, the branching ratio that the both the
$B^{\pm}$ meson decays include a muon. \item $P_{2b/\rm gluon}$,
the angle between $B$ mesons is  narrower than $2 m_{B}/p_{B}$ so
that the two muons  would intersect. \item
$P_{\slashchar{E}_{T}}$, the suppression from the missing $E_{T}$
cut. \item $P_{\rm geo}$, the muons intersect with an angle
between muons of $O(2 m_{a}/p_{B})$.
\end{itemize}
The dominant contribution to the inclusive muon mode of $B^{\pm}$
comes from the mode $B^{\pm}\to D^{0}+ \m +\n +X$ which the
branching ratio $Br\simeq 10^{-1}$. Thus, the background cross
section is suppressed by $Br^{2}=10^{-2}$.

The second suppression factor is estimated by,
\begin{eqnarray}
\label{eq:narrow}
 P_{2b/\rm gluon} \simeq \frac{\sigma_{2b}}{ \sigma_{\rm gluon}}
 \simeq
  \frac{1}{6\pi}
 \int_{4 m_{b}^{2}}^{Q_{\rm Max}^{2}} \frac{dQ^{2}}{Q^{2}}
\a_{s}(Q^{2}) \left(1 + \frac{2 m_{b}^{2}}{Q^{2}} \right) \sqrt{ 1
- \frac{4 m_{b}^{2}}{Q^{2}}},
\end{eqnarray}
where $\sigma_{\rm gluon}\simeq 1\,\m$b is the production cross
section of the two gluon jet for $p_{T}\gtrsim 100$\,GeV. Here, the
upper limit of the integration, $Q^{2}_{Max}$, is  given by,
\begin{eqnarray}
Q_{\rm Max}^{2} = 4 m_{b}^{2} + \frac{1}{4} p_{B}^{2}
\left(2\frac{m_{B}}{p_{B}} \right)^{2},
\end{eqnarray}
which comes from the requirement that the two b-quarks are
collimated in an angle $2 m_{B}/p_{B}$. By using the masses
$m_{B}=5.3$\,GeV and $m_{b}=4.2$\,GeV, we find $P_{2b/\rm gluon}$
is about $10^{-3}$. In this expression, we have assumed that the
gluon multiplicity in a jet is $1$, otherwise the momentum
fraction the $B$ meson pair to the gluon jet gets smaller. If we
take into account the gluon multiplicity, the cross section is
further suppressed (see Ref.\,\cite{Mangano:1992qq} for detailed
discussion of the gluon multiplicity).

The third suppression factor $P_{\slashchar E_{T}}$ is significant
in this background. Similar to the previous background from the
cascade decay, $B\to D+\m+\n$, the momentum fractions of the
muons are suppressed in the inclusive muon modes, since the
dominant contributions to the inclusive muon mode involve the $D$
mesons. For example, if we require that the distribution of the
momentum fraction of the muon pair carrying at least 70\,\% of the
$B$ meson momentum, is less than $10^{-2}$  for each muons with
$p_{B}\gtrsim 50$\,GeV. In addition, since the accompanying jet is a
gluon jet,  the neutrino emission rate in the remaining jets is
much suppressed compared with the $b$-jet case. Thus, we can
expect that the rejection factor from the missing transverse
momentum cut, i.e. $ \slashchar{E}_{T}\le 0.2\, p_{T,\m\m}$, is
much better than $10^{-3}$, i.e. $P_{\slashchar E_{T}}<10^{-3}$.

The suppression factor $P_{\rm geo}$ is much more complicated
compared with the other background events discussed above.
However, since the angle between the two muons must be of $O(2
m_{a}/p_{a})$, while the the muons from the $B$ mesons are emitted
in directions with angles of $O(m_{B}/p_{B})$, the factor $P_{\rm
geo}$ at least involves,
\begin{eqnarray}
 P_{\rm geo}\lesssim  \frac{1}{3^{2}}\left(\frac{m_{a}}{m_{B}} \right)^{2}.
\end{eqnarray}
Here the factor ${1}/{3^{2}}$ comes from the requirement that the
both of muons are emitted in a face-to-face direction so that they
would intersect.

Put the above suppression factors together, the cross section for
the background events are roughly estimated by,
\begin{eqnarray}
 \sigma_{2B\to2\m+X} \,<\, Br_{B^{\pm}\to \m^{\pm}+X}^{2}\times
 P_{\rm geo}\times
 P_{\slashchar E_{T}}\times
 P_{2b/\rm gluon}\times
 \sigma_{\rm gluon},
\end{eqnarray}
and the total rejection factor is expected to be much better than
$10^{-10}$ for $m_{a}\lesssim 1$\,GeV. Therefore, we expect that
the background from the combinatorial muon pair from two $B$ meson
decays is subdominant.

Similar to the $B$ meson background, a combinatorial muon pair
from two $D$ mesons contribute to the background if they
intersect. By replacing $b$-quark and $B$ meson masses with the
$c$-quark and $D$ meson masses in Eq.\,(\ref{eq:narrow}), we
obtain a suppression factor,
\begin{eqnarray}
P_{2D/\rm gluon} \simeq 3\times 10^{-3}.
\end{eqnarray}
$P_{\rm geo}$ should also be modified as
\begin{eqnarray}
 P_{\rm geo}\lesssim  \frac{1}{3^{2}}\left(\frac{m_{a}}{m_{D}} \right)^{2}.
\end{eqnarray}
This lead to a suppression factor of $10^{-9}$ which is a little
bit weaker than that of the $B$ meson. Hoever, the muon isolation
cut is more effective compared to the $B$ meson decay event.
Therefore, again, we expect that the combinatorial background from
the $D$ mesons is also subdominant.

%%%%%%%%%%%%%%%%%%%%%%%%%%%%%%%%%%%%%%%
\subsection{Muon photo-production}\label{sec:photoproduction}

%%%%%%%%%%%%%%%%%%%%%%%%%%%%%%%%%%%%%%%%%%%%%
\begin{table}[tb]
\begin{center}
\begin{tabular}{c|c c c c c}
& $Z$ & $A$ & $\rho$ (g$\cdot$cm$^{-3}$) & $n$ ($10^{22}$cm$^{-3}$)&$P_{\gamma\to\mu\mu}$ \\
\hline
Be & 4 & 9  & 1.85  & 12.3 & 4 $\times 10^{-8}$\\
C & 6 & 12 & 1.9-2.3 &  9.5-11& 7 $\times 10^{-8}$\\
Si & 14 & 28 & 2.33 & 4.98 & 2 $\times 10^{-7}$
\end{tabular}
\end{center}
\caption{The probability for the muon photo-production at the 1\,mm thick material.
In the probability, we have used $C_{\gamma\mu\mu}=3$.
}
\label{tab:material}
\end{table}%
%%%%%%%%%%%%%%%%%%%%%%%%%%%%%%%%%%%%%%%%%%%%%

The photo-production of a muon pair at the detector material is also a serious background
if we look for the R-axion decaying outside of the beam-pipe.
The cross section of such events are given by,
\begin{eqnarray}
 \sigma_{\gamma\to\m\m} = P_{\gamma\to\m\m} \times \sigma_{\gamma},
\end{eqnarray}
where $\sigma_{\gamma}$ is the production cross section of the photon,
$P_{\gamma\to\m\m}$ is a probability of the muon photo-production for
a photon for a given material.

The probability $P_{\gamma\to\m\m}$ is roughly estimated as followings.
The muon photo-production occurs by exchanging a photon in a $t$-channel between
the muons and the target nucleus, and
the cross-section is dominated by the process with the minimum momentum transfer of order of $O(m_{\m}^{2}/k)$ for a given photon momentum $k$.
Thus, we can take the the nucleus of the material as a point-like particle and
neglect the recoil of the nucleus.
In this limit, the differential cross section is approximated by the Bethe-Heitler cross section
(see Ref.\,\cite{Tsai:1973py} for more discussion),
and the total cross section of this process is almost insensitive to the incoming photon momentum;
\begin{eqnarray}
\sigma_{\rm photo-production} = C_{\gamma\mu\mu} \times 10^{-30}{\rm cm}^{2} \times
\left(\frac{Z}{4}\right)^{2}.
\end{eqnarray}
where $Z$ is the charge number of the nucleus.
The small momentum dependence is factorized in a coefficient $C_{\gamma\m\m}=O(1)$.
From this cross section, the photo-production probability per a incoming photon
at the material of thickness ${\mit \Delta}\ell$ is given by
\begin{eqnarray}
 P_{\gamma\to\mu\mu} &=& \sigma_{\rm photo-production} \times n \times {\mit\Delta}\ell,
\end{eqnarray}
where $n$ denotes the number density of the nucleus of the material.
In table\,\ref{tab:material}, we list the photo-production probability
at the 1\,mm thick of Be, C and Si which are  typical material of the beam pipe and the inner detectors.
From the table, we find that the probability of the muon photo-production at 1\,mm thick materials
is $O(10^{-7})$.

\begin{figure}[t]
 \begin{minipage}{.49\linewidth}
  \includegraphics[width=\linewidth]{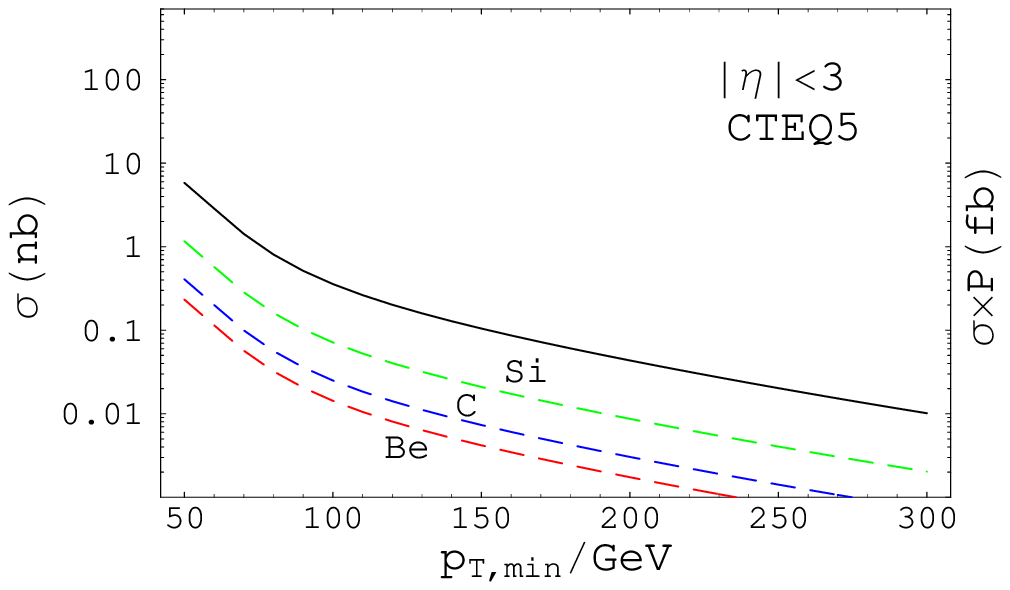}
 \end{minipage}
 \begin{minipage}{.49\linewidth}
  \includegraphics[width=.95\linewidth]{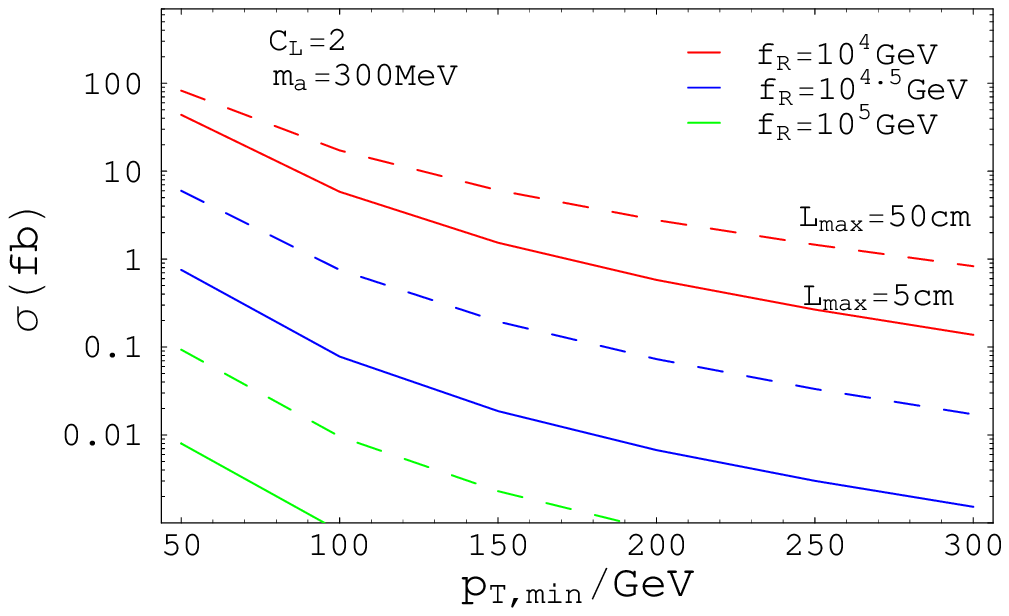}
 \end{minipage}
\caption{Left) The production cross section of direct photon at the LHC with a low $p_{T}$ cutoff (solid line) at leading order.
The dashed lines show the muon photo-production cross section at the 1\,mm thick material,
i.e. $\s\times P$, for Be, C, and Si.
In this figure, we have used $C_{\m\m}=3$ again.
Right) The cross section for the event with the R-axion decaying  within the transverse length
$L_{\rm max}$  for a given low $p_{T}$ cutoff.
The solid  (dashed) lines show the cross section for the event in which the R-axion decays within 5\,cm
(50\,cm) for $f_{R} = 10^{4,4.5,5}$\,GeV from the bottom up.
}
\label{fig:detection}
\end{figure}

Now, let us consider the photon production cross section.
The one of the most important source of the photon at the LHC comes from
the direct photon production via the QCD Compton process,
$qg\to q\gamma$, and the quark annihilation process $qq\to g\gamma$.
In Fig.\,\ref{fig:detection}, we showed the production cross section of the direct photon at the LHC
with a low $p_{T}$ cutoff as a solid line.
The figure shows that the direct photon production cross section is $O(1)$\,nb for $p_{T}\gtrsim 100$\,GeV.
In the same figure, we have also plotted the the muon photo-production cross section at the 1\,mm thick material for Be, C, and Si as dashed lines.

In order to compare the background process with the signature, we also plotted the
cross section for the event with the R-axion decaying  within the transverse length
$L_{\rm max}$  for a given low $p_{T}$ cutoff (solid line).
The result for $L_{\rm max}=5$\,cm corresponds to the R-axion decaying within the beam pipe.
In this case, the main backgrounds from the muon photo-production occur at the
wall of the beam pipes which consist of Be.
By comparing the figures, it is found that that the signal-to-background ratio
is good if we chose $p_{T,min}=100$\,GeV.

For $f_{R} > 10^{4.5}$\,GeV, however, it is difficult to produce enough R-axion for $L_{\rm max} = 5$\,cm
even for the integrated luminosity 100\,fb$^{-1}$.
In this case, we need to extend the region of the displaced vertex search.
For example, if we choose $L_{\rm max} = 50$\,cm, we expect a sizable number of the
R-axion production.
However, in this case, the amount of the material at which the muon photo-production occur
gets much larger.
Therefore, we need to see whether the muon pair candidate for the signal produced
at the detector material or the in between of the material.
For such study, we need more detailed detector simulation.

%%%%%%%%%%%%%%%%%%%%%%%%%%%%%%%%%%%%%%%%%%%%%
%%%%%%%%%%%%%%%%%%%%%%%%%%%%%%%%%%%%%%%%%%%%%

Notice that the photon in jets also contributes to the muon photo-production cross section.
In this case, the cross section for the background process is given by,
 \begin{eqnarray}
 \sigma_{jet\to\mu\mu} = P_{\rm jet}\times P_{\gamma\to \mu\mu}\times \sigma_{\rm jet},
\end{eqnarray}
In the region for $p_{T}=100-500$\,GeV, the cross section of the inclusive
jets are about $7\times 10^{2}$ of the direct photon cross section\,\cite{Giele:1993dj}.
The jet rejection factor comes from the requirement of the momentum balance of the muon pair
and the remaining jets as well as the muon pair isolation cut.
For example,  if we are require $p_{\pi}/p_{\rm jet}>0.7$,
the background is suppressed by $O(10^{-2})$ (see Ref.\,\cite{Shlyapnikov:2001jf}).
Furthermore, the photo-production from the $\pi^{0}$ is always accompanied by
a photon which is not converted into a muon pair.
Therefore, we can expect the jet rejection factor $O(10^{-(2-3)})$, and hence,
the net contribution of the jets to the muon photo-production is
 comparable to the photo-production from the direct photon discussed above.

Finally, we mention here the typical event shape of the muons produced by the photo-production.
The typical invariant mass of the muon pair produced by
the photo-production process is of  $O(2m_{\m})$.
Thus, it seems challenging to tell the signature from
the muons from the photo production by using the invariant mass distribution of the muon pairs.

%%%%%%%%%%%%%%%%%%%%%%%%%%%%%%%%%%%%%%%%%%%%%
\subsection{Summary of the background estimation}\label{sec:summary}
\begin{figure}[t]
 \includegraphics[width=.6\linewidth]{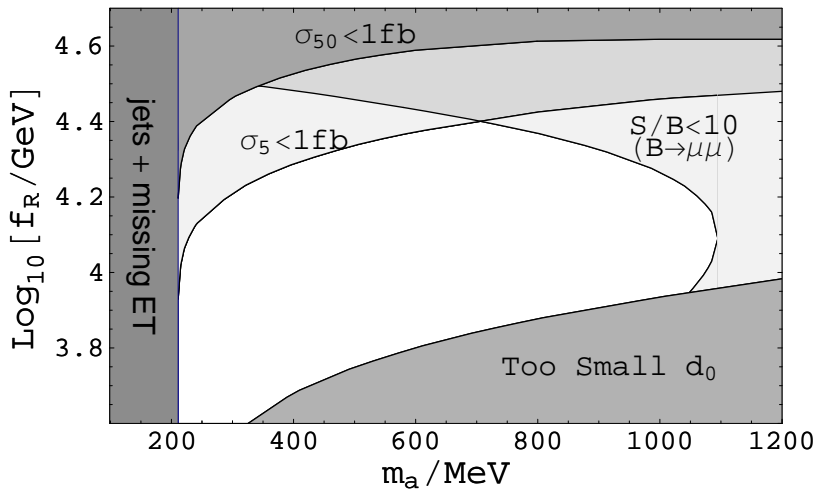}
\caption{
Schematic figure of the detectability of the R-axion at the LHC experiments
for a given values of $m_{a}$ and $f_{R}$ with $C_{L}=2$ fixed.
}
\label{fig:detectability}
\end{figure}

In this section, we gave rough estimations of the backgrounds to the R-axion signature.
As we have shown, the most serious background comes from the inclusive decay
of the $B$ meson, $B^{0}\to D^{\pm}+\m^{\mp}+\n \to \m^{+}\m^{-}+\n\n$,
if we look for the R-axion decaying inside the beam-pipe.
In Fig.~\ref{fig:detectability}, we show a schematic figure of the detectability of the R-axion
at the LHC experiments for a given values of $m_{a}$ and $f_{R}$, with $C_{L}$ fixed at the value
$C_{L}=2$.
In the figure, the shaded regions ``$\sigma_{5}<1\,$fb'' and ``$\sigma_{50}<1\,$fb''
correspond to the parameter spaces where the integrated cross sections of the R-axion
for $L_{T}=1-5\,$cm and $L_{T}=1-50\,$cm are smaller than $1$\,fb, respectively.
In the both regions, we have assumed $p_{T}\ge 100$\,GeV.
Notice that the total cross section is almost independent of the R-axion mass.

In the small $f_{R}$ and the heavy $m_{a}$ region, the lifetime of the R-axion is rather short,
and the muon pairs from the light meson decays can be serious backgrounds.
Even worse, the transverse impact parameter gets too small to be
distinguished from the primary vertex in such region.
We have shown the parameter space where the transverse impact parameter
is too small to be distinguished as the shaded region on the bottom-right corner.

In the figure, we also plotted the region where the signal-to-background ratio is
worse than 10 for the background from the short-lived meson decays,
by assuming that we are looking for the R-axion vertex with
$L_{T}=1-5$\,cm and $p_{T}\ge 100$\,GeV.
In this case the most serious background is the one from 
 $B^{0}\to D^{\pm}+\m^{\mp}+\n \to \m\m+\n\n$.
As we see from the figure, this background is more serious for the heavier R-axion, since the suppression factor $P_{\rm geo}$ in
Eq.\,(\ref{eq:geo}) is less effective for the heavier R-axion.

Notice that the backgrounds from the $B$ meson decays can
be negligible for the longer decay length regions,
for example, for $L_{T}>5$\,cm.
However, in such longer decay length region, the background events
from the $K$ meson decay as well as the  muon photo-production process
at the detector material are significant.
Thus, the R-axion search outside of the beam-pipe requires careful study.

Finally, we comment on the R-axion search for the lighter mass.
When the R-axion mass is lighter than the two times of the muon mass,
the main decay mode of the R-axion is the electron pair mode.
In this case, the R-axion mainly decay outside the detector,
and the R-axion is virtually stable inside the detectors of the LHC experiments.
Thus, the R-axion events just look like jets\,+\,missing $p_{T}$ events.
The detection of the R-axion in such a mass region is quite challenging,
since the background cross section from the jets\,+\,$Z^{0}\to \n\n$ is much larger than the
R-axion production cross section.

%%%%%%%%%%%%%%%%%%%%%%%%%%%%%%%%%%%%%%%%%%%%%
%%%%%%%%%%%%%%%%%%%%%%%%%%%%%%%%%%%%%%%%%%%%%

%%%%%%%%%%%%%%%%%%%%%%%%%%%%%%%%%%%%%%%%%%%%%%%%%
\section{Conclusions}
In this paper, we have discussed the detectability of the R-axion.
The properties of the R-axion are interrelated with
the nature of the supersymmetry breaking sector,
the messenger sector, and the solution of the $\m$ and $B\m$ problems.
Thus, the R-axion can be a powerful probe of the physics behind the SSM.
In the minimal R-breaking scenario, the R-symmetry breaking scale can be as low as $f_{R}\simeq 10^{4-5}$\,GeV, and the R-axion has a mass of  $O(100)$\,MeV.
As we have shown, in such cases, a sizable number of R-axions are produced at the LHC experiments
and can be detected by searching for the displaced vertex left by the R-axion decay.
As a result, we found that the we can detect the R-axion at the LHC experiments
for $m_{a}\simeq 200-1000$\,MeV and $f_{R}\simeq 10^{4}\,$GeV (Fig.\,\ref{fig:detectability}).

In this study, we have assumed that the $\m$-term has a vanishing R-charge.
When we consider models with a $\m$-term which has a non-vanishing R-charge,
the mixing angles of the R-axion and the CP-odd Higgs bosons are different from
the ones given in Sec.\,\ref{sec:rint}.
Especially, when the R-charge of the $\m$ term corresponds to $2$, the mixing through
the $B\m$ term vanishes, and the R-axion does not couple to the Standard Model fermions directly.
In that case, the main decay mode of the R-axion is
into a pair of photons for $m_{a}\lesssim 1\,$GeV
and into a pair of gluons for $m_{a}\gtrsim 1\,$GeV.
In those cases, we need different strategies to search the R-axion.

Several comments are in order.
1) In our analysis, we have only considered the R-axion with $m_{a}=O(100)$\,MeV.
For heavier R-axion, however, this particle can also decay into a pair of the heavier SM fermions.
In that case, the R-axion does not leave a displaced vertex, and hence, we need
another strategy to find the R-axion.
2) The R-axion searches  in the rare decays of such as $\Upsilon(1S)$ and
$J/\psi$ particles are also interesting possibilities.
So far, the constraints on $f_{R}$ from those rare decay modes are given $f_{R}\gtrsim 10^{3}$\,GeV
coming from $Br(\Upsilon\to \g+a)<10^{-(5-6)}$ for $m_{a}\lesssim 1$\,GeV\,\cite{:2008hs}
(see also Ref.\,\cite{Kim:1986ax} for a review of the experimental constraints on the axion-like particle).%
\footnote{The study whether the R-axion can explain the HyperCP observation\,\cite{Park:2005eka}
may be also interesting\,\cite{He:2006fr}.}
3) Since the R-axion couples to heavy fermions such as top-quark and the gauginos
rather strongly, the production process accompanied by those heavy fermions might be
interesting channels.
4) At the LEP experiments, the dominant production cross section
comes from $e^+e^- \to Z (\gamma)\to Z (\gamma) + a $ via the anomaly coupling
given in Eq.\,(\ref{eq:Ranom})
and $e^+e^- \to Z\to h(H) + a$ via the mixing between R-axion and CP-odd Higgs bosons.
For $f_R\gtrsim 10^4$\,GeV, we found that the cross sections of those processes are below fb, and hence,
we had no chance to have produced the R-axion at the LEP experiments.
5) At the Tevatron experiments, we need to search for a low $p_T$  R-axion for the integrated luminosity
$10$\,fb$^{-1}$ (see Fig.\,\ref{fig:production}), which requires careful study.%
\footnote{In Ref.\,\cite{Adams:2008yw}, no displaced vertex has been found at Tevatron
by searching for a displaced vertex in the range of 5\,cm$< L_{T}<$20\,cm
by using a cut on the impact parameter $d_{0}>100\,\m$m.
Our signal discussed above would be excluded by the cut in this study
in some parameter space.
It is interesting to see how Tevatron can do in searching for the R-axion in our parameter space.
}

\section*{Acknowledgements}
We appreciate M.~Peskin for a lot of discussion and advice very
much. We also appreciate T.~Barklow and D.~Miller for useful
comments.
MI also appreciate D.~Su for useful discussion.
MI appreciate Y.~Nakayama and T.T~Yanagida for useful discussion 
on the low energy properties of the R-axion.
HSG would also like to thank I.~Hinchliffe, M.~Shapiro,
J.~Thaler and D.~Walker for discussions. We appreciate the
hospitality of the. Aspen Center for Physics, where this
collaboration began. The work of MI was supported by the U.S.
Department of Energy under contract number DE-AC02-76SF00515. The
work of HSG was supported in part by DOE under contract number
DE-AC02-05CH11232 and by the U.S. National Science Foundation
under grants PHY-04-57315.

\end{document}